\documentclass[a4paper]{article}
\usepackage{jheppub}
\usepackage[utf8]{inputenc}
\usepackage{amsmath,amsthm,amssymb,mathtools,tikz,booktabs,tensor,slashed,dutchcal,upgreek}
\usepackage[mathscr]{eucal}
\usetikzlibrary{cd}

\newcommand{\mc}{\mathcal} \newcommand{\mf}{\mathfrak} \newcommand{\mb}{\mathbb} \newcommand{\on}{\operatorname}  \newcommand{\ms}{\mathscr}
\newcommand{\slot}{\;\cdot\;} \newcommand{\la}{\langle} \newcommand{\ra}{\rangle}
\newcommand{\gm}{{\mc G}} \newcommand{\di}{\slashed D} \newcommand{\tr}{\on{Tr}}
\DeclareMathOperator{\gr}{\mathcal{R}}
\title{Direct derivation of $\ms N=1$ supergravity in ten dimensions to all orders in fermions}
\author[a]{Julian Kupka,}
\emailAdd{j.kupka@herts.ac.uk}
\author[a]{Charles Strickland-Constable,}
\emailAdd{c.strickland-constable@herts.ac.uk}
\author[a]{and Fridrich Valach}
\emailAdd{f.valach@herts.ac.uk}
\affiliation[a]{Department of Physics, Astronomy and Mathematics,
University of Hertfordshire, College Lane, Hatfield, AL10 9AB, United Kingdom}
\abstract{It has been known for some time that generalised geometry provides a particularly elegant rewriting of the action and symmetries of 10-dimensional supergravity theories, up to the lowest nontrivial order in fermions. By exhibiting the full symmetry calculations in the second-order formalism, we show in the $\ms N=1$ case that this analysis can be upgraded to all orders in fermions and we obtain a strikingly simple form of the action as well as of the supersymmetry transformations, featuring overall only five higher-fermionic terms. Surprisingly, even after expressing the action in terms of classical (non-generalised geometric) variables one obtains a simplification of the usual formulae. This in particular confirms that generalised geometry provides the natural set of variables for studying (the massless level of) string theory. We also show how this new reformulation implies the compatibility of the Poisson--Lie T-duality with the equations of motion of the full supergravity theory.}

\usepackage{thmtools}

\allowdisplaybreaks

\begin{document}
  \maketitle
  \section{Introduction and conclusion}
    $\ms N=1$ supergravity in ten dimensions plays an important role in string theory, since (for a particular choice of the gauge group) it describes the two-derivative part of the massless sector of heterotic and type I superstring theories. Despite the fact that the explicit form of this theory has been known for many decades now~\cite{Bergshoeff:1981um,Chapline:1982ww,Dine:1985rz} (see also~\cite{Bergshoeff:1988nn,Bergshoeff:1989de} for a slightly simpler treatment), 
    and significant geometric structure has been found to underlie the theory at lowest nontrivial order in fermions~\cite{Siegel:1993th,CSW1,Garcia-Fernandez:2013gja,Coimbra:2014qaa}, there has been little progress in a similar understanding the structure of the higher fermion terms, which are crucial for the theory to be truly supersymmetric.\footnote{Notably, some progress has been achieved in \cite{Jeon:2011sq,Cederwall:2016ukd,Butter:2021dtu} in the context of double field theory, using superspace and other techniques.}

    In this paper we present a direct derivation of $\ms N=1$ supergravity coupled to Yang--Mills multiplets, precisely by formulating the question in terms of 
    the geometric approach of generalised geometry.
    This results in the following surprisingly simple form of the action:
    \begin{equation}\label{eq:a}
        S=\int_M\!\!\mc R\sigma^2+\bar\psi_{\alpha}\slashed D\psi^{\alpha}+\bar\rho\slashed D\rho+2\bar\rho D_{\alpha}\psi^{\alpha}-\tfrac1{768}\sigma^{-2}(\bar\psi_{\alpha}\gamma_{cde}\psi^{\alpha})(\bar\rho\gamma^{cde}\rho)-\tfrac1{384}\sigma^{-2}(\bar\psi_{\alpha}\gamma_{cde}\psi^{\alpha})(\bar\psi_{\beta}\gamma^{cde}\psi^{\beta}).
    \end{equation}
    which is invariant under the supersymmetry transformations
    \begin{equation}\label{eq:s}
    \begin{aligned}
        \delta \mc G_{ab}&=\delta\mc G_{\alpha\beta}=0,\quad \delta\mc G_{a\beta}=\delta\mc G_{\beta a}=\tfrac12\sigma^{-2}\bar \epsilon\gamma_a\psi_{\beta}\\
        \delta\sigma&=\tfrac18\sigma^{-1}(\bar\rho\epsilon)\\
        \delta\rho&=\slashed D\epsilon+\tfrac1{192}\sigma^{-2}(\bar\psi_{\alpha}\gamma_{cde}\psi^{\alpha})\gamma^{cde}\epsilon\\
        \delta\psi_{\alpha}&=D_{\alpha}\epsilon+\tfrac18\sigma^{-2}(\bar\psi_{\alpha}\rho)\epsilon+\tfrac18\sigma^{-2}(\bar\psi_{\alpha}\gamma_c\epsilon)\gamma^c\rho
    \end{aligned}
    \end{equation}
    The relevant notation and details are discussed in Section \ref{sec:gg}. Decomposing the expressions in terms of the standard fields via
    \[\gm,\sigma\quad \leadsto\quad \text{metric }g,\text{ Kalb-Ramond field }B,\text{ gauge field }A,\text{ dilaton }\varphi\]
    \[\rho \quad \leadsto\quad \text{dilatino }\uprho\qquad\qquad \psi \quad \leadsto\quad \text{gravitino }\uppsi,\text{ gaugino }\upchi\]
    we arrive at the action
  \begin{align}
    S=\!\smash{\int_M}\!&\Phi(R+4|\nabla \varphi|^2-\tfrac1{12}H_{\mu\nu\rho}H^{\mu\nu\rho}+\tfrac14\tr F_{\mu\nu}F^{\mu\nu}-\bar\uppsi^\mu\slashed\nabla\uppsi_\mu+\uprho\slashed\nabla\uprho+\tfrac12\tr\bar\upchi\slashed\nabla_{\hspace{-1mm}A}\upchi-2\bar\uppsi^\mu\nabla_\mu\uprho\nonumber\\
    &\;\quad+\tfrac14\bar\uppsi^\mu\slashed H\uppsi_\mu-\tfrac14\bar\uprho\slashed H\uprho-\tfrac18\tr\bar\upchi\slashed H\upchi+\tfrac12 H_{\mu\nu\rho}\bar\uppsi^\mu \gamma^\nu\uppsi^\rho+\tfrac14\bar\uppsi^\mu H_{\mu\nu\rho}\gamma^{\nu\rho}\uprho \label{eq:origa}\\
    &\;\quad+\tfrac12\tr\bar\upchi\slashed F\uprho+\tr F_{\mu\nu}\bar\uppsi^\mu \gamma^\nu\upchi+\tfrac1{384}(\bar\uppsi_{\mu}\gamma_{\nu\rho\sigma}\uppsi^{\mu})(\bar\uprho\gamma^{\nu\rho\sigma}\uprho)-\tfrac1{768}(\bar\uprho\gamma^{\mu\nu\rho}\uprho)\tr(\bar\upchi\gamma_{\mu\nu\rho}\upchi)\nonumber\\
    &-\tfrac1{192}(\bar\uppsi_{\mu}\gamma_{\rho\sigma\tau}\uppsi^{\mu})(\bar\uppsi_{\nu}\gamma^{\rho\sigma\tau}\uppsi^{\nu})+\tfrac1{192}(\bar\uppsi_{\mu}\gamma_{\nu\rho\sigma}\uppsi^{\mu})\tr(\bar\upchi\gamma^{\nu\rho\sigma}\upchi)-\tfrac1{768}\tr(\bar\upchi\gamma_{\mu\nu\rho}\upchi)\tr(\bar\upchi\gamma^{\mu\nu\rho}\upchi)),\nonumber
  \end{align}
  with $\Phi:=\sqrt{|g|}e^{-2\varphi}$ and the supersymmetry transformations
    \begin{equation}\label{eq:origs}
    \begin{aligned}
      \delta g_{\mu\nu}&=\bar\upepsilon\gamma_{(\mu}\uppsi_{\nu)}\\
      \delta B_{\mu\nu}&=\bar\upepsilon\gamma_{[\mu}\uppsi_{\nu]}-\tr A_{[\mu}\bar\upepsilon\gamma_{\nu]}\upchi\\
      \delta A_\mu&=-\tfrac12\bar\upepsilon\gamma_\mu\upchi\\
      \delta \varphi&=\tfrac14\bar\uprho\upepsilon-\tfrac14\bar\uppsi^\mu\gamma_\mu\upepsilon\\
      \delta \uprho&=-\slashed\nabla \upepsilon+(\nabla_\mu\varphi)\gamma^\mu\upepsilon+\tfrac14\slashed H\upepsilon+\tfrac1{96}(\bar\uppsi_\mu \gamma_{\nu\rho\sigma}\uppsi^\mu)\gamma^{\nu\rho\sigma}\upepsilon+\tfrac14(\bar\uprho\upepsilon)\uprho-\tfrac1{192}\tr(\bar\upchi\gamma_{\mu\nu\rho}\upchi)\gamma^{\mu\nu\rho}\upepsilon\\
      \delta\uppsi_\mu&=\nabla_\mu\upepsilon-\tfrac18H_{\mu\nu\rho}\gamma^{\nu\rho}\upepsilon-\tfrac14(\bar\uppsi_\mu\uprho)\upepsilon-\tfrac14(\bar\uppsi_\mu\gamma_\nu\upepsilon)\gamma^\nu\uprho+\tfrac14(\bar\uprho\upepsilon)\uppsi_\mu\\
      \delta\upchi&=\tfrac12\slashed F\upepsilon-\tfrac14(\bar\upchi\uprho)\upepsilon-\tfrac14(\bar\upchi\gamma_\mu\upepsilon)\gamma^\mu\uprho+\tfrac14(\bar\uprho\upepsilon)\upchi,
    \end{aligned}
    \end{equation}
    which --- although significantly longer than the preceding expressions --- still provide a simplification compared to the standard treatment. In the above we followed the usual conventions \eqref{conv}, \eqref{conva}, together with $\slashed C:=\tfrac1{p!}C_{\mu_1\dots \mu_p}\gamma^{\mu_1\dots \mu_p}$ for a $p$-form $C$.\footnote{Note that our definition of the dilatino follows \cite{CSW1} (up to a sign). Setting $\uplambda=\gamma_\mu\uppsi^\mu+\uprho$ we obtain the more standard definition, which leads to the simple transformation $\delta\varphi=\tfrac14\bar\uplambda\upepsilon$ but is less natural from the viewpoint of generalised geometry. Note also that in the variation of $\uppsi$ the cubic fermionic terms all contain $\uppsi$, $\uprho$, $\upepsilon$ but in different combinations. In principle one could use Fierz identities to obtain a more homogeneous-looking expression; however, the price to pay for this would be the appearance of more complicated numerical coefficients. Same applies to $\delta\upchi$.}

    The simplicity of the action \eqref{eq:a} and of the transformations \eqref{eq:s} allows one to check the local supersymmetry of the action by hand --- we exhibit the entire calculation (as well as the equations of motion) in Section \ref{sec:theory}. Due to the uniqueness of the supersymmetric extensions we know that the action \eqref{eq:origa} and supersymmetry \eqref{eq:origs} have to coincide with the originally found form \cite{Bergshoeff:1981um,Chapline:1982ww,Dine:1985rz}, up to simple field redefinitions (and Fierz identities).
    
    We note that the formulation \eqref{eq:a}, \eqref{eq:s} is fully geometric and only requires very mild topological assumptions (namely that the bundle $C_+$ given by the generalised metric admits a spin structure). It also obviates the use of supercovariant derivatives, commonly employed in the usual approach. Finally, the generalised-geometric formulation makes manifest the compatibility of the supergravity equations with Poisson--Lie T-duality \cite{Klimcik:1995ux} (see Section \ref{sec:pltd} for details).
    
    The more detailed structure of the paper is as follows. In Section \ref{sec:gg} we provide an introduction to generalised geometry, with the more technical details and derivations moved to Appendices \ref{app:gg} and \ref{app:unpack}. In Section \ref{sec:theory} we recall the details of the proposed theory, display its equations of motion, discuss generalisations of the setup, and finally provide a full derivation of the local supersymmetry of the action to all orders. The last Section \ref{sec:pltd} discusses Poisson--Lie T-duality and its compatibility with the supergravity equations of motion and with the local supersymmetry transformations. The spinor conventions and the full list of the required Fierz identities can be found in Appendix \ref{app:fierz}.
    
  \section{Generalised geometry} \label{sec:gg}
    In this section we provide an introduction to the necessary aspects of generalised geometry: the theory of Courant algebroids. We start by briefly mentioning the more formal approach and then provide the specific details for the case at hand.
    \subsection{Courant algebroids}
      The central notion of generalised geometry is that of a \emph{Courant algebroid} \cite{liu1997manin,let}. This is a mathematical structure which (roughly speaking) captures the symmetries of the massless sector of string theory. More concretely, a Courant algebroid consists of the following data:
      \begin{itemize}
        \item a (smooth) vector bundle $E\to M$
        \item an $\mb R$-bilinear bracket on the space of sections $\Gamma(E)$ of $E$
        \item a non-degenerate symmetric bilinear pairing $\langle \slot,\slot\rangle$ on the fibers of $E$
        \item a vector bundle map $a\colon E\to TM$ called the \emph{anchor}
      \end{itemize}
      which satisfies the following axioms (for all $u,v,w\in\Gamma(E)$, $f\in C^\infty(M)$):
      \begin{itemize}
        \item Jacobi identity\footnote{In the literature this is also commonly called the Leibniz identity, with the name Jacobi identity used for the vanishing of the sum of cyclic permutations of $[[u,v],w]$, which is how the corresponding axiom for Lie algebras is typically written. However, it can be argued that already in the Lie algebra context, the more natural way of writing the axiom is $\on{ad}_u[v,w]=[\text{ad}_uv,w]+[v,\text{ad}_uw]$, which is the reason for our choice of nomenclature.} $[u,[v,w]]=[[u,v],w]+[v,[u,w]]$
        \item the derivation property for multiplication by functions $[u,fv]=f[u,v]+(a(u)f)v$
        \item the invariance of inner product $a(u)\langle v,w\rangle=\langle [u,v],w\rangle+\langle v,[u,w]\rangle$
        \item the symmetric part of the bracket is governed by the pairing $[u,v]+[v,u]=\ms D\langle u,v\rangle$,
      \end{itemize}
      where we have defined $\ms D\colon C^\infty(M)\to\Gamma(E)$ by $\langle u,\ms Df\rangle:=a(u)f$. Throughout the remainder of this paper we will also use the identification $E\cong E^*$ provided by $\langle \slot,\slot\rangle$.
      
      From these axioms one can prove other useful formulae, such as
      \[a([u,v])=[a(u),a(v)]\]
      or $a\circ a^*=0$, which can be equivalently stated as the fact that
      \[0\to T^*M\xrightarrow{a^*}E\xrightarrow aTM\to 0\]
      is a chain complex. When the latter is in fact an exact sequence, the algebroid is called \emph{exact}. More generally, if $a$ is surjective it is called \emph{transitive}.
      
      From the axioms it also follows that for any $u\in\Gamma(E)$ one can define the \emph{generalised Lie derivative} $\ms L_u$, which can act on any section of $E^{\otimes n}$ or the tensor product of any such section with any (half-)density on $M$. For instance, for $v\in \Gamma(E)$ and $\tau$ a (half-)density on $M$, we have
      \[\ms L_uv=[u,v],\qquad \ms L_u\tau=L_{a(u)}\tau,\]
      where $L$ is the ordinary Lie derivative. This is then extended using the Leibniz rule.
      
      Transitive Courant algebroids can be classified locally \cite{let}. The result is that \emph{locally} any transitive Courant algebroid over $M$ looks like the ``trivial'' model
      \begin{equation}\label{local}
        TM\oplus T^*M\oplus (\mf g\times M),
      \end{equation}
      where the trivial vector bundle in the last summand is constructed using a Lie algebra $\mf g$ with an invariant nondegenerate symmetric bilinear form (or \emph{quadratic Lie algebra} for short), which we will denote simply by $\tr$. Thus the sections of $E$ consists of a formal sum of a vector field, 1-form field, and a $\mf g$-valued function. The remaining structures are given by
      \begin{equation}\label{bkt}
        \begin{gathered}
          a(x+\alpha+s):=x,\qquad \langle x+\alpha+s,y+\beta+t\rangle:=\alpha(y)+\beta(x)+\tr st\\
          [x+\alpha+s,y+\beta+t]:=L_xy+(L_x\beta-i_yd\alpha+\tr t\,ds)+(L_xt-L_ys+[s,t]_\mf g).
        \end{gathered}
      \end{equation}
      This bracket encodes the gauge structure of the relevant supergravity \cite{Coimbra:2014qaa}. Note that in order for the kinetic term for the gauge fields to have the correct sign, one requires the bilinear form $\tr$ to be negative-definite.
      Globally, one can use Courant algebroid automorphisms to glue this local description over different patches in $M$ to a global one.
      
      To see a global example, suppose we start with a principal $G$-bundle over $M$, with a connection $\nabla_{\!A}$ with curvature $F$ and a 3-form $H\in\Omega^3(M)$, satisfying
      \[dH=\tfrac12\tr (F\wedge F).\]
      Then we obtain a transitive Courant algebroid
      $E=TM\oplus T^*M\oplus \on{ad}_G$, where the last summand corresponds to the associated adjoint vector bundle. The anchor and inner product take the same form as for the above local model, while one can write the bracket concisely as \cite{Coimbra:2014qaa}
      \begin{align*}
        [x+\alpha+s,y+\beta+t]&=L_xy+(L_x\beta-i_yd\alpha+\tr[(\nabla_{\!A}s)t-(i_xF)t+(i_yF)s]+i_yi_xH)\\
        &\qquad\qquad\qquad +(i_x\nabla_{\!A}t-i_y\nabla_{\!A}s+[s,t]_\mf g+i_yi_xF).
      \end{align*}
      
      Due to our focus on 10-dimensional $\ms N=1$ supergravity we will (through most of the text) restrict our attention to transitive Courant algebroids with $\dim M=10$; though it will be clear from the calculations that many of the results apply directly to the more general setups.
    
    \subsection{Bosonic fields}\label{subsec:bos}
      Let us now turn to the supergravity fields, starting with the generalised metric. This is defined simply as a map of vector bundles $\gm\colon E\to E$ which is both symmetric and satisfies $\gm^2=\on{id}$. Such a map induces an orthogonal decomposition $E= C_+\oplus C_-$ into $\pm 1$ eigenbundles; in turn, $\gm$ can be completely reconstructed from $C_+$. In the present case, we shall focus on the cases where
      \begin{itemize}
        \item the anchor $a$ restricted to $C_+$ is an isomorphism
        \item the induced inner product $\langle \slot,\slot\rangle|_{C_+}$ has signature $(9,1)$
        \item $C_+$ admits a spin structure.
      \end{itemize}
      The last (very mild) condition is there simply to ensure the existence of the appropriate spinor bundles for the description of the fermionic fields. The first two conditions ensure that we recover the usual physical fields. More concretely, assuming $\dim M=10$ and following \cite{Coimbra:2014qaa}, any $C_+$ in \eqref{local} which satisfies these conditions has the form
      \begin{equation}\label{cplusparam}
        C_+=\{x+(i_xg+i_xB-\tfrac12\tr A \,i_xA)+i_xA\mid x\in TM\}
      \end{equation}
      for some Lorentzian metric $g$, Kalb--Ramond field $B\in\Omega^2(M)$, and gauge field $A\in\Omega^1(M,\mf g)$. (We will soon see how the dilaton enters.) We also record here that it follows that
      \begin{equation}\label{cminus}
        \begin{aligned}
        C_-&=C'_-\oplus C''_-\\
        C'_-&:=\{x+(-i_xg+i_xB-\tfrac12\tr A\,i_xA)+i_xA\mid x\in TM\}\\
        C''_-&:=\{0-\tr tA+t\mid t\in \mf g\times M\},
        \end{aligned}
      \end{equation}
      with all the subbundles $C_+$, $C_-'$, $C_-''$ orthogonal to each other.
      To understand the role of the $\tfrac12\tr (A\,i_xA)$ term, note that in the parametrisation \eqref{cplusparam} the anchor map gives an isometry (up to a numerical factor $\pm2$) between $(C_\pm,\langle\slot,\slot\rangle|_{C_\pm})$ and $(TM,g)$, i.e.
      \begin{equation}\label{notisometry}
        \langle u_\pm,u_\pm\rangle=\pm2g(a(u_\pm),a(u_\pm)),\qquad \forall u_+\in C_+, \; u_-\in C_-'.
      \end{equation}
    
      To encode the dilaton, let now $H$ be the line bundle of half-densities on $M$. To have a concrete picture in mind note that any choice of coordinate system $x^\mu$ on $M$ gives rise to a local section $\sqrt{|dx^1\wedge\dots\wedge dx^{\dim M}|}$ of $H$; changing the coordinates corresponds to multiplying this section with $|\text{Jacobian}|^{-1/2}$. The product of two half-densities gives a density and can thus be naturally integrated over $M$ (even when $M$ is not orientable).
      
      The bosonic field content of our theory will be encoded via a generalised metric $\gm$ and an everywhere non-vanishing half-density $\sigma\in \Gamma(H)$. To recover the standard description of the dilaton in terms of a scalar function $\varphi$ one writes
      \begin{equation}\label{eq:dil}
        \sigma^2=\Phi=\sqrt{|g|}e^{-2\varphi},
      \end{equation}
      where $\sqrt{|g|}$ is the metric density. 
      Note that our description here is equivalent to realising the bosonic fields as a $G$-structure, with $G$ the stabiliser of the generalised metric inside $O(p,q)$, for the generalised frame bundle with enhanced structure group $O(p,q)\times\mathbb{R}^+$ as in~\cite{CSW1}. 
      
      At several points in the present text it will be convenient to work with an explicit frame compatible with $\gm$. For this purpose we will always use a local frame \[e_A=\{e_a,e_\alpha\}\] which is adjusted to the decomposition $E=C_+\oplus C_-$,\footnote{i.e.\ $e_a$ is a frame of $C_+$ and $e_\alpha$ is a frame of $C_-$} and which satisfies $\langle e_A,e_B\rangle=\eta_A\delta_{AB}$ for some $\eta_A\in\{\pm1\}$. We will call such a frame \emph{orthonormal}. One of the big advantages of such a frame is the fact that the structure coefficients $c_{ABC}$ of the Courant algebroid, defined by
      \[c_{ABC}:=\langle [e_A,e_B],e_C\ra,\]
      are completely antisymmetric.
      
    \subsection{Fermionic fields}
      Under the above assumptions on $E$ and $C_+$ (in particular concerning the signature of the latter), let us denote the vector bundles of positive and negative chirality Majorana--Weyl spinors associated to $C_+$ by $S_\pm$. The fermionic fields of our model are then
      \[\rho\in \Gamma(\Pi S_+\otimes H),\qquad \psi\in\Gamma(\Pi S_-\otimes C_-\otimes H),\]
      where $\Pi$ stands for the parity shift (i.e.\ it states that the fields are taken to be anticommuting). Note that $\psi$ has a $C_-$-valued vector index and a spinor index w.r.t.\ the spinor bundle for $C_+$. This distinction, which follows \cite{CSW1}, is absolutely crucial and heavily restricts the possible terms in the action and variations which one can write down. 
      Also note that we have defined $\rho$ and $\psi$ to be half-densities, following the insights from~\cite{Kupka:2024tic} (see also \cite{Bergshoeff:1988nn}). This results in further simplifications to the form of the action and variations below. 
      
      Let us again look concretely at what this reproduces in the case \eqref{local}, \eqref{bkt} with the generalised metric given by \eqref{cplusparam}. Under the identifications
      \[C_+\cong TM,\qquad C_-=C_-'\oplus C_-''\cong TM\oplus (\mf g\times M)\]
      we decompose $\rho$ and $\psi$ as
      \[\rho=\sqrt[4]2\sigma\uprho,\qquad \psi=\sqrt[4]2\sigma\uppsi+\tfrac1{\sqrt[4]2}\sigma\upchi,\]
      where we included some factors of $\sqrt[4]2$ for convenience. The fields in the decomposition are
      \begin{itemize}
        \item the dilatino $\uprho$, i.e.\ a positive chirality Majorana spinor w.r.t.\ the Lorentzian metric $g$
        \item the gravitino $\uppsi$, i.e.\ a negative chirality Majorana vector-spinor
        \item the gaugino $\upchi$, i.e.\ a Lie algebra-valued negative chirality Majorana spinor.
      \end{itemize}
    
    \subsection{Generalised connections}\label{subsec:con}
      In this Subsection we follow \cite{CSW1}. For any Courant algebroid we can define \emph{generalised connections} as natural generalisations of affine connections on $TM$. A generalised connection is thus an $\mb R$-bilinear map 
      \[D\colon \Gamma(E)\times\Gamma(E)\to\Gamma(E),\quad (u,v)\mapsto D_uv,\]
      such that $D_{fu}=fD_u$, $D_u(fv)=fD_uv+(a(u)f)v$, and $D\langle \slot,\slot\rangle=0$. In the last formula we have again extended $D_u$ (via the product rule) to act on any tensors in $E$.
      Any generalised connection $D$ acts naturally also on densities and half-densities via
      \[D_u\mu:=\ms L_u\mu-\mu D_Au^A,\qquad D_u\sigma:=\ms L_u\sigma-\tfrac12\sigma D_Au^A,\]
      with $\mu$ and $\sigma$ a density and a half-density, respectively.
      For any orthonormal frame we also define the \emph{connection coefficients} $\Gamma_{ABC}$ by
      \[(D_{e_A}v)^B=a(e_A)v^B+\Gamma_{A}{}^B{}_C v^C.\]
      Due to the last condition in the definition of a generalised connection these satisfy \[\Gamma_{ABC}=-\Gamma_{ACB}.\]
      
      We will say that a generalised connection $D$ is \emph{Levi-Civita for $\gm$ and compatible with $\sigma$}, denoted simply by $D\in LC(\gm,\sigma)$, if the following holds:
      \begin{itemize}
        \item $D\gm=0$, or equivalently $\Gamma_{Aa\alpha}=\Gamma_{A\alpha a}=0$,
        \item $D\sigma=0$, or equivalently $\Gamma^{A}{}_{AB}=\on{div} e_B$,
        \item $D$ is torsion-free, i.e.\ $\Gamma_{[ABC]}=-\tfrac13c_{ABC}$,
      \end{itemize}
    where \[\on{div} u:=\sigma^{-2}\ms L_u\sigma^2,\quad u\in\Gamma(E)\] is the divergence w.r.t.\ $\sigma$. Note that in particular this implies
    \[D_Au^A=\on{div}u,\qquad \forall u\in\Gamma(E).\]

      Crucially, $LC(\gm,\sigma)$ is nonempty, though typically quite large \cite{Garcia-Fernandez:2016ofz}. In other words, the choice of bosonic fields does not determine uniquely or naturally some specific Levi-Civita connection. However, there are several ``bits'' of $D$ which are uniquely fixed by $\gm$ and $\sigma$ \cite{Siegel:1993th,CSW1,Coimbra:2014qaa,Garcia-Fernandez:2016ofz} --- and these are precisely the ones that are required for our supergravity description. For instance, it follows from the above constraints that
      \[\Gamma_{a\beta\gamma}=-c_{a\beta\gamma},\qquad \Gamma_{\alpha bc}=-c_{\alpha bc},\]
      which is equivalent to the statement
      \[D_{u_+}v_-=[u_+,v_-]_-,\qquad D_{u_-}v_+=[u_-,v_+]_+,\]
      where subscripts $\pm$ denote the orthogonal projections $E\to C_\pm$. Another such operator is the \emph{Dirac operator} \[\di:=\gamma^aD_a,\] acting on spinor half-densities (w.r.t.\ $C_+$), e.g.\ the $\rho$ field. To see this, note that a choice of an orthonormal frame produces a local trivialisation of our vector bundles, so that in particular \[\Gamma(S_+\otimes H)\cong \Gamma(H)\otimes S_+^0,\] where $S_+^0$ is the vector space of positive Majorana spinors. In this identification we have
      \begin{equation}\label{diracloc}
        \di \rho=\gamma^a\ms L_{e_a}\rho+\tfrac14\Gamma_{abc}\gamma^{abc}\rho=\gamma^a\ms L_{e_a}\rho-\tfrac1{12}c_{abc}\gamma^{abc}\rho.
      \end{equation}
      where $\ms L$ is now understood as only acting on the half-density part of the expression. Other uniquely-defined operators are
        \begin{alignat}{3}
          \di\psi^{\alpha }&=\gamma^a\ms L_{e_a}\psi^{\alpha }+\tfrac14\Gamma_{abc}\gamma^{abc}\psi^{\alpha }+\Gamma_{a}{}^{\alpha }{}_{\beta }\gamma^a\psi^{\beta }&&=\gamma^a\ms L_{e_a}\psi^{\alpha }-\tfrac1{12}c_{abc}\gamma^{abc}\psi^{\alpha }-c_{a}{}^{\alpha }{}_{\beta }\gamma^a\psi^{\beta },\label{diracloc2}\\
          D_{\alpha }\rho&=\ms L_{e_{\alpha }}\rho+\tfrac14\Gamma_{\alpha bc}\gamma^{bc}\rho-\tfrac12\Gamma^{\gamma }{}_{\gamma \alpha }\rho&&=\ms L_{e_{\alpha }}\rho-\tfrac14c_{\alpha bc}\gamma^{bc}\rho-\tfrac12(\on{div} e_{\alpha })\rho,\label{derivloc}\\
          D_{\alpha }\psi^{\alpha }&=\ms L_{e_{\alpha }}\psi^{\alpha }+\tfrac14\Gamma_{\alpha bc}\gamma^{bc}\psi^{\alpha }+\tfrac12\Gamma^{\gamma }{}_{\gamma \alpha }\psi^{\alpha }&&=\ms L_{e_{\alpha }}\psi^{\alpha }-\tfrac14c_{\alpha bc}\gamma^{bc}\psi^{\alpha }+\tfrac12(\on{div} e_{\alpha })\psi^{\alpha }.\nonumber
        \end{alignat}
         We see that in all these the dependence on the representative in $LC(\gm,\sigma)$ vanishes, and in addition $\di \rho$ and $\di \psi^\alpha$ are also independent of $\sigma$. In particular the kinetic terms $\bar\psi_{\alpha }\slashed D\psi^{\alpha }$, $\bar\rho\slashed D\rho$, and $\bar\rho D_{\alpha }\psi^{\alpha }$ appearing below depend only on $(\psi,\gm)$, $(\rho,\gm)$, and $(\rho,\psi,\gm,\sigma)$, respectively.
    
    As a final useful fact, note that for any generalised connection we have (assuming a compact support) that for any $u\in \Gamma(E)$ and $\mu$ a density
    \[\int_M D_A(u^A\mu)=\int_M (D^A u_A)\mu+D_u\mu=\int_M \ms L_u\mu=0,\]
    allowing us to use integration by parts.
    
    \subsection{Curvature operators}\label{subsec:curv}
      Curvature tensors in generalised geometry have been introduced in numerous works \cite{Siegel:1993th,CSW1,Hohm:2011si,Garcia-Fernandez:2013gja,Coimbra:2014qaa,Severa:2016lwc,Garcia-Fernandez:2016ofz,Jurco:2016emw,sv2}. Here we provide a brief review of the concepts and identities relevant for the task at hand.
      
      For any $D\in LC(\gm,\sigma)$ we can construct the \emph{generalised Riemann tensor} $\gr_{ABCD}$ as
      \begin{equation}\label{riemann}
        \gr(w,z,x,y):=\tfrac12w^D y^B(x^A[D_A,D_B]z_D+z^A[D_A,D_D]x_B-(D_A x_B) (D^A z_D)).
      \end{equation}
      One can check that this is indeed a tensor, and has the following symmetries:
      \begin{equation}\label{riemsym}
        \gr_{ABCD}=\gr_{[AB]CD}=\gr_{AB[CD]}=\gr_{CDAB},\qquad \gr_{A[BCD]}=0.
      \end{equation}
      Unfortunately, the generalised Riemann tensor depends on the choice of the representative $D$ in $LC(\gm,\sigma)$. However, one can construct a \emph{generalised Ricci tensor} $\gr_{AB}$ and \emph{generalised scalar curvature} $\gr$ which do not, by
      \[\gr_{a\gamma}=\gr_{\gamma a}:=4\gr^b{}_{ab\gamma},\quad \gr_{ac}=\gr_{\alpha\gamma}=0,\qquad \gr:=2\gr^{ab}{}_{ab}.\]
    For instance, in an orthonormal frame we have \cite{Siegel:1993th,sv2,Streets:2024rfo}
    \begin{equation}\label{scalarexplicit}
      \gr=-2(\on{div} e^a)(\on{div} e_a)-4a(e^a)\on{div} e_a+\tfrac13c_{abc}c^{abc}+c_{ab\gamma}c^{ab\gamma}.
    \end{equation}
    
    The unimportant prefactors $2$ and $4$ in the definition of $\gr_{AB}$ and $\gr$ are related to the conventions which we have adopted in the present work.\footnote{Changing the definitions in order to get rid of these factors would introduce unwanted prefactors elsewhere (the \emph{whack-a-mole} principle).}
    To justify the particular choice of contractions, note that we have the identities \eqref{zigzag}, \eqref{contractions}, \eqref{contractions2}:
    \[\gr_{a\beta c\delta}=0, \qquad \gr^{b}{}_{ab\gamma}=\gr^{\beta}{}_{a\beta\gamma},\qquad \gr^{ab}{}_{ab}+\gr^{\alpha\beta}{}_{\alpha\beta}=\Psi,\]
    where $\Psi\in C^\infty(M)$ is independent of $D$, $\gm$, $\sigma$ (it is intrinsic to the Courant algebroid structure).\footnote{This in particular shows that the present definition of the Ricci tensor coincides with the one in \cite{Streets:2024rfo} (in the case of half-densities), while the scalar curvature differs from the one in \cite{Streets:2024rfo} by a function independent of $\gm$ and $\sigma$. Note that the formulas in \cite{Streets:2024rfo} possess a certain symmetry in regard to $C_+$ and $C_-$; the fact that in the present text we require the less symmetric definition is related to the asymmetric nature of the $\ms N=1$ supergravity.}
    
    Finally, we note the generalisations of the usual formulas linking the Dirac operator with the curvatures. These are the \emph{Lichnerowicz formula}
        \begin{equation}\label{lich}
          (\di^2+D^{\alpha}D_{\alpha})\epsilon=-\tfrac18\mc R\epsilon
        \end{equation}
    and the formula
        \begin{equation}\label{eq:lichric}
          [\di, D_{\beta}]\epsilon=\tfrac14\gr_{a\beta}\gamma^a\epsilon,
        \end{equation}
    with $\epsilon\in\Gamma(\Pi S_-\otimes H)$, generalising the ones in \cite{CSW1}. These are proven in appendices \ref{app:lich} and \ref{app:lichric}, respectively.
          
  \section{The theory and its local supersymmetry}\label{sec:theory}
    \subsection{The field content, action, and supersymmetry transformations}
      Let us now summarise all the ingredients of the theory.
      
      While the physical spacetime remains a ten-dimensional manifold $M$, we consider an enhanced physical background given by a transitive Courant algebroid $E$ with base $M$. 
      Over this background, we then have the field content of the theory as follows:
      \begin{itemize}
        \item a generalised metric \[\mc G\in\on{End}(E),\] satisfying $\mc G^T=\mc G$ and $\mc G^2=1$, corresponding to an orthogonal splitting $E=C_+\oplus C_-$ s.t.
        \begin{itemize}
          \item $C_+$ has signature $(9,1)$ and admits spinors
          \item the anchor map $a$ gives an isomorphisms between $C_+$ and $TM$
        \end{itemize}
        \item an everywhere non-vanishing half-density $\sigma$
        \item a generalised dilatino $\rho\in\Gamma(\Pi S_+\otimes H)$
        \item a generalised gravitino $\psi\in\Gamma(\Pi S_-\otimes C_-\otimes H)$.
      \end{itemize}
      Finally, the supersymmetry parameter is \[\epsilon\in\Gamma(\Pi S_-\otimes H).\]
      As before, $H$ and $S_\pm$ denote the half-density line bundle (w.r.t.\ $M$) and the Majorana--Weyl spinor bundles for $C_+$, respectively.
      We claim that in terms of these variables the action of $\ms N=1$ supergravity coupled to Yang--Mills multiplets is 
      \[
        S=\int_M\!\!\mc R\sigma^2+\bar\psi_{\alpha}\slashed D\psi^{\alpha}+\bar\rho\slashed D\rho+2\bar\rho D_{\alpha}\psi^{\alpha}-\tfrac1{768}\sigma^{-2}(\bar\psi_{\alpha}\gamma_{cde}\psi^{\alpha})(\bar\rho\gamma^{cde}\rho)-\tfrac1{384}\sigma^{-2}(\bar\psi_{\alpha}\gamma_{cde}\psi^{\alpha})(\bar\psi_{\beta}\gamma^{cde}\psi^{\beta})
      \]
      and the supersymmetry variations are
      \begin{equation*}
    \begin{aligned}
        \delta \mc G_{ab}&=\delta\mc G_{\alpha\beta}=0,\quad \delta\mc G_{a\beta}=\delta\mc G_{\beta a}=\tfrac12\sigma^{-2}\bar \epsilon\gamma_a\psi_{\beta}\\
        \delta\sigma&=\tfrac18\sigma^{-1}(\bar\rho\epsilon)\\
        \delta\rho&=\slashed D\epsilon+\tfrac1{192}\sigma^{-2}(\bar\psi_{\alpha}\gamma_{cde}\psi^{\alpha})\gamma^{cde}\epsilon\\
        \delta\psi_{\alpha}&=D_{\alpha}\epsilon+\tfrac18\sigma^{-2}(\bar\psi_{\alpha}\rho)\epsilon+\tfrac18\sigma^{-2}(\bar\psi_{\alpha}\gamma_c\epsilon)\gamma^c\rho
    \end{aligned}
    \end{equation*}     
    Recall that we use the indices $a,b,c,\dots$, $\alpha,\beta,\gamma,\dots$ (and $A,B,C,\dots$) for $C_+$, $C_-$ (and $E$) respectively.
    As discussed in the Introduction, the above claim follows from the facts that
    \begin{itemize}
      \item the action is invariant under said supersymmetry transformations (this is shown in the remainder of this section)
      \item the action and supersymmetry transformations reduce to \eqref{eq:origa} and \eqref{eq:origs}, respectively (this is shown in appendix \ref{app:unpack})
      \item up to quadratic order in fermions the expressions \eqref{eq:origa} and \eqref{eq:origs} coincide with the standard ones \cite{Bergshoeff:1981um,Chapline:1982ww,Dine:1985rz} (and thus also the generalised-geometric expressions from \cite{CSW1,Coimbra:2014qaa})
    \end{itemize}
    and from the uniqueness of the supergravity action.
  Note that the two quartic terms appearing in the action are in fact the only ones (up to Fierz identities) compatible with the generalised-geometric index structure.
  
  Finally, for later reference we also include the equations of motion, obtained using formulas from appendix \ref{subsec:vars}:
  \begin{equation}\label{eqs}
  \begin{aligned}
    0&=\gr_{a\alpha}+\,\sigma^{-2}(\tfrac12\bar\psi_\beta\gamma_aD_\alpha\psi^\beta+\bar\psi_\alpha\gamma_aD_\beta\psi^\beta-\bar\psi_\beta\gamma_aD^\beta\psi_\alpha+\tfrac12\bar\rho\gamma_aD_\alpha\rho-\tfrac12\bar\psi_\alpha D_a\rho\\
    &\qquad\qquad\qquad\qquad+\tfrac14\bar\rho\gamma_{ab}D^b\psi_\alpha-\tfrac14\bar\psi_\alpha\gamma_{ab}D^b\rho),\\
    0&=\gr+\,\sigma^{-2}(2\bar\psi^\alpha D_\alpha \rho+2\bar\rho D_\alpha \psi^\alpha)+\sigma^{-4}[\tfrac1{768}(\bar\psi_\alpha\gamma_{cde}\psi^\alpha)(\bar\rho\gamma^{cde}\rho)\\
    &\qquad\qquad\qquad\qquad+\tfrac1{384}(\bar\psi_\alpha\gamma_{cde}\psi^\alpha)(\bar\psi_\beta\gamma^{cde}\psi^\beta)],\\
    0&=\di\rho+D_\alpha\psi^\alpha-\tfrac1{768}\sigma^{-2}(\bar\psi_\alpha\gamma_{cde}\psi^\alpha)\gamma^{cde}\rho,\\
    0&=\di\psi^\alpha-D^\alpha\rho-\sigma^{-2}[\tfrac1{768}(\bar\rho\gamma_{cde}\rho)\gamma^{cde}\psi^\alpha+\tfrac1{192}(\bar\psi_\beta\gamma_{cde}\psi^\beta)\gamma^{cde}\psi^\alpha].
  \end{aligned}
  \end{equation}
    
    \subsubsection{A generalisation}\label{subsec:generalisation}
      Let us now consider the more general context in which we take an arbitrary Courant algebroid $E\to M$ (without imposing transitivity or the condition $\dim M=10$) together with an arbitrary generalised metric $\gm\in \on{End}(E)$ satisfying $\gm^T=\gm$ and $\gm^2=\on{id}$, such that $\on{rank}C_-\neq1$, the subbundle $C_+$ admits spinors and has signature either $(9,1)$, $(5,5)$, or $(1,9)$. The latter requirement is needed for the existence of Majorana--Weyl spinors and for the Fierz identities to hold. The condition on the rank of $C_-$ is required for the space $LC(\gm,\sigma)$ to be non-empty (cf.\ \cite{Streets:2024rfo}, see also Appendix \ref{app:diff}).
            Other fields, as well as the supersymmetry parameter, remain sections of the same vector bundles as in the preceding subsection. 
            
      The main point here is that the action \eqref{eq:a} is still invariant under the supersymmetry transformations \eqref{eq:s} and leads to the equations of motion \eqref{eqs}. Of course, it can no longer be reduced down to yield the usual supergravity \eqref{eq:origa}, \eqref{eq:origs}. Nevertheless, this generalisation is quite useful for several reasons. First, it is needed for showing the compatibility of supergravity with the Poisson--Lie T-duality (see Section \ref{sec:pltd}). Second, by taking various special cases one recovers theories which can either serve as useful toy models or lead to theories which are interesting in their own right.
      
      For instance, in the special case $\gm=\on{id}$ (corresponding to $E=C_+$) one recovers the dilatonic supergravity theory of \cite{Kupka:2024tic}. This is a topological theory, whose field content only consists of the dilaton and dilatino.
      
      Another interesting limit is obtained by taking the manifold $M$ to be a point. In this case the field space becomes finite-dimensional and all the expressions become purely algebraic. Nevertheless, the theory is still symmetric under \eqref{eq:s}, and so it provides a convenient toy model for understanding the structure of the fully physical setup.
      
    \subsection{Invariance under local supersymmetry}
    The supersymmetry variation of the action is
    \begin{align*}
      \delta S&=\int_M[-\tfrac12\gr^{a\alpha }\bar\psi_{\alpha }\gamma_a\epsilon+\tfrac14\gr\bar\rho\epsilon]\\
      &\vphantom{\di^2}+[2\bar\psi^{\alpha }\di D_{\alpha }\epsilon+\tfrac14\sigma^{-2}(\bar\psi_{\alpha }\rho)(\bar\epsilon\di\psi^{\alpha })-\tfrac14\sigma^{-2}(\bar\psi_{\alpha }\gamma_a\epsilon)(\bar\rho\gamma^a\di\psi^{\alpha })\\
      &\quad-\tfrac12\sigma^{-2}(\bar\psi_{\gamma }\gamma_a\epsilon)(\tfrac12\bar\psi_{\alpha }\gamma^aD^{\gamma }\psi^{\alpha }+\bar\psi^{\gamma }\gamma^aD^{\alpha }\psi_{\alpha }-\bar\psi_{\alpha }\gamma^aD^{\alpha }\psi^{\gamma })]\\
      &+[2\bar\rho\di^2\epsilon+\tfrac1{96}\sigma^{-2}(\bar\psi^{\alpha }\gamma_{(3)}\psi_{\alpha })(\bar\epsilon\gamma^{(3)}\di\rho)-\tfrac14\sigma^{-2}(\bar\psi^{\alpha }\gamma_a\epsilon)(\bar\rho\gamma^aD_{\alpha }\rho)]\\
      &+[\vphantom{\di^2}2\bar\rho D_{\alpha }D^{\alpha }\epsilon-\tfrac14\sigma^{-2}(\bar\psi_{\alpha }\rho)(\bar\epsilon D^{\alpha }\rho)+\tfrac14\sigma^{-2}(\bar\psi_{\alpha }\gamma_a\epsilon)(\bar\rho\gamma^aD^{\alpha }\rho)-2\bar\psi^{\alpha }D_{\alpha }\di\epsilon\\
      &\quad\vphantom{\di^2}+\tfrac1{96}\sigma^{-2}(\bar\psi^{\gamma }\gamma_{(3)}\psi_{\gamma })(\bar\epsilon\gamma^{(3)}D_{\alpha }\psi^{\alpha })+\sigma^{-2}(\bar\psi_{\alpha }\gamma_a\epsilon)(-\tfrac12\bar\psi^{\alpha }D^a\rho+\tfrac14\bar\rho\gamma^{ab}D_b\psi^{\alpha }-\tfrac14\bar\psi^{\alpha }\gamma^{ab}D_b\rho)\\
      &\quad\vphantom{\di^2}-\tfrac14\sigma^{-2}(\bar\rho\epsilon)(\bar\psi^{\alpha }D_{\alpha }\rho+\bar\rho D_{\alpha }\psi^{\alpha })]\\
      &\vphantom{\di^2}+[\tfrac1{3072}\sigma^{-4}(\bar\rho\epsilon)(\bar\psi^{\alpha }\gamma_{(3)}\psi_{\alpha })(\bar\rho\gamma^{(3)}\rho)\\
      &\quad\vphantom{\di^2}-\tfrac1{384}\sigma^{-2}\bar\psi^{\alpha }\gamma_{(3)}(D_{\alpha }\epsilon+\tfrac18\sigma^{-2}(\bar\psi_{\alpha }\rho)\epsilon+\tfrac18\sigma^{-2}(\bar\psi_{\alpha }\gamma_a\epsilon)\gamma^a\rho)(\bar\rho\gamma^{(3)}\rho)\\
      &\quad\vphantom{\di_i}-\tfrac1{384}\sigma^{-2}(\bar\psi^{\alpha }\gamma_{(3)}\psi_{\alpha })\bar\rho\gamma^{(3)}(\di \epsilon+\tfrac1{192}\sigma^{-2}(\bar\psi^{\gamma }\gamma'_{(3)}\psi_{\gamma })\gamma'^{(3)}\epsilon)]\\
      &\vphantom{\di_i}+[\tfrac1{1536}\sigma^{-4}(\bar\rho\epsilon)(\bar\psi^{\alpha }\gamma_{(3)}\psi_{\alpha })(\bar\psi^{\gamma }\gamma^{(3)}\psi_{\gamma })\\
      &\quad\vphantom{\di_i}-\tfrac1{96}\sigma^{-2}(\bar\psi^{\alpha }\gamma_{(3)}\psi_{\alpha })\bar\psi^{\gamma }\gamma^{(3)}(D_{\gamma }\epsilon+\tfrac18\sigma^{-2}(\bar\psi_{\gamma }\rho)\epsilon+\tfrac18\sigma^{-2}(\bar\psi_{\gamma }\gamma_a\epsilon)\gamma^a\rho)]
    \end{align*}
    
    \subsubsection{Quadratic order}
      As the first step in showing $\delta S=0$ we consider the terms quadratic in fermionic variables ($\rho$, $\psi$, and $\epsilon$).
      First, the terms containing $\rho$ and $\epsilon$ combine to
      \[(\delta S)_{\rho\epsilon}=\int_M\tfrac14\mc R(\bar\rho\epsilon)+2\bar\rho \di^2\epsilon+2\bar\rho D_{\alpha }D^{\alpha }\epsilon=0\]
      due to the Lichnerowicz formula \eqref{lich}.
      Similarly,
      \[(\delta S)_{\psi\epsilon}=\int_M -\tfrac12\gr^{a\alpha }\bar \psi_{\alpha }\gamma_a\epsilon+2\bar\psi^{\alpha }\di D_{\alpha }\epsilon-2\bar\psi^{\alpha }D_{\alpha }\di\epsilon=\int_M 2\bar \psi^{\alpha }(-\tfrac14\gr_{a\alpha }\gamma^a+[\di, D_{\alpha }])\epsilon,\]
      which again vanishes due to \eqref{eq:lichric}.
    
    \subsubsection{Quartic order}
      Using integration by parts and \eqref{eq:f1} we calculate
      \begin{align*}
        (\delta S)_{\psi\psi\psi\epsilon}&=\int_M\sigma^{-2}\left[-\tfrac14(\bar\psi_{\alpha }\gamma^a D_{\gamma }\psi^{\alpha })\bar\psi^{\gamma }\gamma_a-\tfrac12(\bar\psi^{\alpha }\gamma^aD_{\gamma }\psi^{\gamma })\bar\psi_{\alpha }\gamma_a+\tfrac12(\bar\psi_{\gamma }\gamma^a D^{\gamma }\psi^{\alpha })\bar\psi_{\alpha }\gamma_a\right.\\
        &\vphantom{\int_M}\qquad\left.+\tfrac1{96}(\bar\psi^{\alpha }\gamma_{(3)}\psi_{\alpha })D_{\gamma }\bar\psi^{\gamma }\gamma^{(3)}+\tfrac1{48}(\bar\psi^{\alpha }\gamma_{(3)}D_{\gamma }\psi_{\alpha })\bar\psi^{\gamma }\gamma^{(3)}+\tfrac1{96}(\bar\psi^{\alpha }\gamma_{(3)}\psi_{\alpha })D_{\gamma }\bar\psi^{\gamma }\gamma^{(3)}\right]\epsilon\\
        &=\int_M\sigma^{-2}[(\tfrac1{96}+\tfrac1{96}-\tfrac1{48})(\bar\psi^{\alpha }\gamma_{(3)}\psi_{\alpha })(D_{\gamma }\bar\psi^{\gamma }\gamma^{(3)}\epsilon)+(-\tfrac1{48}+\tfrac1{48})(D_{\gamma }\bar\psi_{\alpha }\gamma_{(3)}\psi^{\alpha })(\bar\psi^{\gamma }\gamma^{(3)}\epsilon)]\\
        &\vphantom{\int_m}=0.
      \end{align*}
      Similarly,
        \begin{align*}
          (\delta S)_{\psi\rho\rho\epsilon}&=\int_M\sigma^{-2}[-\tfrac14(\bar\rho\gamma^a D_{\alpha }\rho)\bar\psi^{\alpha }\gamma_a-\tfrac14(\bar\psi_{\alpha }\rho)D^{\alpha }\bar\rho+\tfrac14(\bar\rho\gamma^aD^{\alpha }\rho)\bar\psi_{\alpha }\gamma_a-\tfrac14(\bar\psi^{\alpha }D_{\alpha }\rho)\bar\rho\\
          &\vphantom{\int_M}\qquad-\tfrac14(\bar\rho D_{\alpha }\psi^{\alpha })\bar\rho+\tfrac1{384}(\bar\rho\gamma^{(3)}\rho)D_{\alpha }\bar\psi^{\alpha }\gamma_{(3)}+\tfrac1{192}(\bar\rho\gamma^{(3)}D_{\alpha }\rho)\bar\psi^{\alpha }\gamma_{(3)}]\epsilon\\
          &\stackrel{\eqref{eq:f2}}{=}\int_M\sigma^{-2}[-\tfrac14(\bar\psi_{\alpha }\rho)D^{\alpha }\bar\rho-\tfrac14(\bar\psi^{\alpha }D_{\alpha }\rho)\bar\rho+\tfrac1{192}(\bar\rho\gamma^{(3)}D_{\alpha }\rho)\bar\psi^{\alpha }\gamma_{(3)}]\epsilon\stackrel{\eqref{eq:fc1}}{=}0.
        \end{align*}
      Finally, to show the vanishing of
      \begin{align*}
        (\delta S)_{\psi\psi\rho\epsilon}&=\int_M\sigma^{-2}[\tfrac14(\bar\psi_{\alpha }\rho)\overline{\di\psi^{\alpha }}-\tfrac14(\bar\rho\gamma^a\di\psi^{\alpha })\bar\psi_{\alpha }\gamma_a+\tfrac1{96}(\bar\psi^{\alpha }\gamma_{(3)}\psi_{\alpha })\overline{\di\rho}\gamma^{(3)}\\
        &\vphantom{\int_M}\qquad+(-\tfrac12\bar\psi^{\alpha }D^a\rho+\tfrac14\bar\rho\gamma^{ab}D_b\psi^{\alpha }-\tfrac14\bar\psi^{\alpha }\gamma^{ab}D_b\rho)\bar\psi_{\alpha }\gamma_a\\
        &\vphantom{\int_M}\qquad+\tfrac1{192}(\bar\psi^{\alpha }\gamma_{(3)}D_a\psi_{\alpha })\bar\rho\gamma^{(3)}\gamma^a+\tfrac1{384}(\bar\psi^{\alpha }\gamma_{(3)}\psi_{\alpha })D_a\bar\rho\gamma^{(3)}\gamma^a]\epsilon
      \end{align*}
      we first show the vanishing of the above terms containing $D\rho$:
      \begin{align*}
        &\int_M\sigma^{-2}[\tfrac1{96}(\bar\psi^{\alpha }\gamma_{(3)}\psi_{\alpha })\overline{\di\rho}\gamma^{(3)}-\tfrac12(\bar\psi^{\alpha }D^a\rho)\bar\psi_{\alpha }\gamma_a-\tfrac14(\bar\psi^{\alpha }\gamma^{ab}D_b\rho)\bar\psi_{\alpha }\gamma_a+\tfrac1{384}(\bar\psi^{\alpha }\gamma_{(3)}\psi_{\alpha })D_a\bar\rho\gamma^{(3)}\gamma^a]\epsilon\\
        &\stackrel{\eqref{eq:f2},\eqref{eq:f21}}{=}\int_M \sigma^{-2}(\bar\psi^{\alpha }\gamma_{bcd}\psi_{\alpha })[-\tfrac1{96}D_a\bar\rho\gamma^a\gamma^{bcd}-\tfrac12\tfrac1{96}D^a\bar\rho\gamma^{bcd}\gamma_a-\tfrac14(-\tfrac1{32}D_a\bar\rho\gamma^{abcd}-\tfrac5{32}D^b\bar\rho\gamma^{cd})\\
        &\qquad\vphantom{\int_M}+\tfrac1{384}D_a\bar\rho\gamma^{bcd}\gamma^a]\epsilon\\
        &=\int_M\sigma^{-2}(\bar\psi^{\alpha }\gamma_{bcd}\psi_{\alpha })D^a\bar\rho(-\tfrac1{96}\gamma_a\gamma^{bcd}-\tfrac1{384}\gamma^{bcd}\gamma_a+\tfrac1{128}\gamma_a{}^{bcd}+\tfrac5{128}\delta_a^{b}\gamma^{c d})\epsilon=0.
      \end{align*}
      Plugging this back and using gamma contractions we get
        \begin{align*}
          (\delta S)_{\psi\psi\rho\epsilon}&=\int_M\sigma^{-2}[\tfrac14(\bar\psi_{\alpha }\rho)\overline{\di\psi^{\alpha }}-\tfrac14(\bar\rho\gamma^a\di\psi^{\alpha })\bar\psi_{\alpha }\gamma_a+\tfrac14(\bar\rho\gamma^{ab}D_b\psi^{\alpha })\bar\psi_{\alpha }\gamma_a\\
          &\vphantom{\int_M}\qquad+\tfrac1{192}(\bar\psi^{\alpha }\gamma_{(3)}D_a\psi_{\alpha })\bar\rho\gamma^{(3)}\gamma^a]\epsilon\\
          &= \int_M \sigma^{-2}[-\tfrac14 (\bar\psi_{\alpha} \rho) D_a \bar\psi^{\alpha}\gamma^a +(-\tfrac14 + \tfrac14) (\bar \rho \gamma^{a b} D_b \psi^{\alpha}) \bar\psi_{\alpha} \gamma_a - \tfrac14 (\bar\rho D_a \psi^{\alpha }) \bar\psi_{\alpha } \gamma^a\\
          &\vphantom{\int_M}\qquad+\tfrac1{192}(\bar\psi^{\alpha }\gamma_{(3)}D_a\psi_{\alpha })\bar\rho\gamma^{(3)}\gamma^a]\epsilon \\
          &\stackrel{\eqref{eq:fc1}}= \int_M \sigma^{-2} (-\tfrac1{192}+\tfrac1{192})(\bar\psi^{\alpha }\gamma_{(3)}D_a\psi_{\alpha })(\bar\rho\gamma^{(3)}\gamma^a\epsilon)=0.
        \end{align*}    
      Since there are no $(\delta S)_{\rho\rho\rho\epsilon}$ terms this concludes the invariance of the action up to terms quartic in fermions.
      
    \subsubsection{Sextic order}
      Finally, we have to show the vanishing of the two sextic terms, $(\delta S)_{\psi\psi\psi\psi\rho\epsilon}$ and $(\delta S)_{\psi\psi\rho\rho\rho\epsilon}$.
      For the latter one we set
      \[\Xi:=(\bar\psi^{\alpha }\gamma_{(3)}\psi_{\alpha })(\bar\rho\gamma^{(3)}\rho)(\bar\rho\epsilon)\]
      and then using
      \[(\bar\rho\gamma^{bcd}\rho)(\bar\psi^{\alpha }\gamma_{bcd}\gamma^a\rho)=-(\bar\rho\gamma^{bcd}\rho)(\bar\psi^{\alpha }\gamma^a\gamma_{bcd}\rho)+6(\bar\rho\gamma^{acd}\rho)(\bar\psi^{\alpha }\gamma_{cd}\rho)\stackrel{\eqref{eq:3r2},\eqref{eq:3r3}}=0\]
      we calculate
      \begin{align*}
        (\delta S)_{\psi\psi\rho\rho\rho\epsilon}&=\int_M\sigma^{-4}[(\bar\rho\gamma^{(3)}\rho)(\bar\psi_{\alpha }\rho)(\bar\psi^{\alpha }\gamma_{(3)}\epsilon)-\Xi]\\
        &\stackrel{\eqref{eq:f2}}=\int_M\sigma^{-4}[\tfrac1{96}(\bar\rho\gamma^{(3)}\rho)(\bar\psi_{\alpha }\gamma_{(3)}'\psi^{\alpha })(\bar\rho\gamma'^{(3)}\gamma_{(3)}\epsilon)-\Xi]\\
        &\stackrel{\eqref{eq:3r3}}=\int_M\sigma^{-4}[\tfrac1{96}(\bar\rho\gamma^{(3)}\rho)(\bar\psi_{\alpha }\gamma_{(3)}'\psi^{\alpha })(\bar\rho\{\gamma'^{(3)},\gamma_{(3)}\}\epsilon)-\Xi]\\
        &=\int_M\sigma^{-4}[\tfrac9{48}(\bar\rho\gamma^{abc}\rho)(\bar\psi_{\alpha }\gamma_{aef}\psi^{\alpha })(\bar\rho\gamma^{ef}{}_{bc}\epsilon)-\tfrac98\Xi]\\
        &=\int_M\sigma^{-4}[\tfrac9{48}(\bar\rho\gamma^{abc}\rho)(\bar\psi_{\alpha }\gamma_{aef}\psi^{\alpha })(\bar\rho(\gamma_{bc}\gamma^{ef}+4\delta^e_b\gamma_c{}^f+2\delta^{ef}_{bc})\epsilon)-\tfrac98\Xi]\\
        &\stackrel{\eqref{eq:3r2}}=9\int_M\sigma^{-4}[\tfrac1{12}(\bar\rho\gamma^{abc}\rho)(\bar\psi_{\alpha }\gamma_{abf}\psi^{\alpha })(\bar\rho\gamma_c{}^f\epsilon)-\tfrac1{12}\Xi]\\
        &=\tfrac34\int_M\sigma^{-4}[(\bar\rho\gamma^{abc}\rho)(\bar\psi_{\alpha }\gamma_{abf}\psi^{\alpha })(\bar\rho(\gamma_c\gamma^f-\delta^f_c)\epsilon)-\Xi]\\
        &\stackrel{\eqref{eq:fcc}}=\tfrac34\int_M\sigma^{-4}\Xi(2-1-1)=0.
      \end{align*}
      
    Using the symmetry in the exchange of $abc$ with $def$ in $(\bar\psi^{\alpha }\gamma_{abc}\psi_{\alpha })(\bar\psi^{\gamma }\gamma_{def}\psi_{\gamma })$ we then have for the remaining term
    \begin{align*}
      (\delta S)_{\psi\psi\psi\psi\rho\epsilon}&=\tfrac1{768}\int_M\sigma^{-4}[-\tfrac1{96}(\bar\psi^{\alpha }\gamma_{(3)}\psi_{\alpha })(\bar\psi^{\gamma }\gamma'_{(3)}\psi_{\gamma })(\bar\rho\gamma^{(3)}\gamma'^{(3)}\epsilon)+\tfrac12(\bar\psi^{\alpha }\gamma_{(3)}\psi_{\alpha })(\bar\psi^{\gamma }\gamma^{(3)}\psi_{\gamma })(\bar\rho\epsilon)\\
      &\vphantom{\int_M}\qquad\qquad\qquad\qquad-(\bar\psi^{\alpha }\gamma_{(3)}\psi_{\alpha })(\bar\psi_{\gamma }\rho)(\bar\psi^{\gamma }\gamma^{(3)}\epsilon)-(\bar\psi^{\alpha }\gamma_{(3)}\psi_{\alpha })(\bar\psi^{\gamma }\gamma^{(3)}\gamma^a\rho)(\bar\psi_{\gamma }\gamma_a\epsilon)]\\
      &\stackrel{\eqref{eq:f2}}=\tfrac1{768}\int_M\sigma^{-4}[-\tfrac1{48}(\bar\psi^{\alpha }\gamma_{(3)}\psi_{\alpha })(\bar\psi^{\gamma }\gamma'_{(3)}\psi_{\gamma })(\bar\rho\gamma^{(3)}\gamma'^{(3)}\epsilon)+\tfrac12(\bar\psi^{\alpha }\gamma_{(3)}\psi_{\alpha })(\bar\psi^{\gamma }\gamma^{(3)}\psi_{\gamma })(\bar\rho\epsilon)\\
      &\vphantom{\int_M}\qquad\qquad\qquad\qquad-(\bar\psi^{\alpha }\gamma_{(3)}\psi_{\alpha })(\bar\psi^{\gamma }\gamma^{(3)}\gamma^a\rho)(\bar\psi_{\gamma }\gamma_a\epsilon)]\\
      &\stackrel{\eqref{eq:f3}}=\tfrac1{768}\int_M\sigma^{-4}(\bar\psi^{\alpha }\gamma_{abc}\psi_{\alpha })[-\tfrac1{96}(\bar\psi^{\gamma }\gamma^{def}\psi_{\gamma })(\bar\rho\{\gamma^{abc},\gamma_{def}\}\epsilon)+\tfrac12(\bar\psi^{\gamma }\gamma^{abc}\psi_{\gamma })(\bar\rho\epsilon)\\
      &\vphantom{\int_M}\qquad\qquad\qquad\qquad-\tfrac58(\bar\psi^{\gamma }\gamma^{abc}\psi_{\gamma })(\bar\rho\epsilon)-\tfrac3{16}(\bar\psi^{\gamma }\gamma^{abc}\gamma_{de}\psi_{\gamma })(\bar\rho\gamma^{de}\epsilon)\\
      &\vphantom{\int_M}\qquad\qquad\qquad\qquad-\tfrac1{192}(\bar\psi^{\gamma }\gamma^{abc}\gamma_{defg}\psi_{\gamma })(\bar\rho\gamma^{defg}\epsilon)]\\
      &=\tfrac1{768}\int_M\sigma^{-4}(\bar\psi^{\alpha }\gamma_{abc}\psi_{\alpha })[-\tfrac1{96}(\bar\psi^{\gamma }\gamma^{def}\psi_{\gamma })\bar\rho(18\delta^a_d\gamma^{bc}{}_{ef}-12\delta^{abc}_{def})\epsilon-\tfrac18(\bar\psi^{\gamma }\gamma^{abc}\psi_{\gamma })(\bar\rho\epsilon)\\
      &\vphantom{\int_M}\qquad\qquad\qquad-\tfrac3{16}\bar\psi^{\gamma }(6\delta^a_d\gamma^{bc}{}_e)\psi_{\gamma }(\bar\rho\gamma^{de}\epsilon)-\tfrac1{192}\bar\psi^{\gamma }(\gamma^{abc}{}_{defg}-36\delta^{ab}_{de}\gamma^c{}_{fg})\psi_{\gamma }(\bar\rho\gamma^{defg}\epsilon)]\\
      &=\tfrac1{768}\int_M\sigma^{-4}(\bar\psi^{\alpha }\gamma_{abc}\psi_{\alpha })[(\bar\psi^{\gamma }\gamma^{aef}\psi_{\gamma })(\bar\rho\gamma^{bc}{}_{ef}\epsilon)(-\tfrac3{16}+\tfrac3{16})+(\bar\psi^{\gamma }\gamma^{abc}\psi_{\gamma })(\bar\rho\epsilon)(\tfrac18-\tfrac18)\\
      &\vphantom{\int_M}\qquad\qquad\qquad-\tfrac1{192}(\bar\psi^{\gamma }\gamma^{abc}{}_{defg}\psi_{\gamma })(\bar\rho\gamma^{defg}\epsilon)]\\
      &\stackrel{\eqref{eq:flip}}=-\tfrac1{36864}\int_M\sigma^{-4}(\bar\psi^{\alpha }\gamma_{abc}\psi_{\alpha })(\bar\psi^{\gamma }\gamma_{def}\psi_{\gamma })(\bar\rho\gamma^{abcdef}\epsilon)=0.
    \end{align*}
    This concludes the proof of the supersymmetry invariance of the action to all orders in fermions.

  \section{Compatibility with the Poisson--Lie T-duality}\label{sec:pltd}
    \subsection{Courant algebroid pullbacks}
    Poisson--Lie T-duality \cite{Klimcik:1995ux} in the context of supergravity can be elegantly stated using the language of Courant algebroids, in the following way \cite{Severa:2016lwc,sv2} (for a double-field theoretic approach see \cite{Hassler:2017yza,Butter:2023nxm}).
    Suppose we have a pullback of vector bundles along a surjective submersion $\pi$:
    \begin{equation}\label{pullback}
      \begin{tikzcd}
      E'\ar[r]\ar[d]&E\ar[d]\\
      M'\ar[r,"\pi"]&M
    \end{tikzcd}
    \end{equation}
    and suppose that both of these are equipped with a Courant algebroid structure such that for all $u,v\in \Gamma(E)$
    \[\pi^*[u,v]=[\pi^*u,\pi^*v],\qquad \pi^*\la u,v\ra=\la \pi^*u,\pi^*v\ra,\qquad \pi_*a(\pi^*u)=a(u).\]
    We then call $E'$ a \emph{Courant algebroid pullback} of $E$. Note that for a given Courant algebroid $E$ and a map $M'\to M$ the Courant algebroid pullback does not need to exist nor be unique. The possible Courant algebroid pullbacks were characterised in \cite{li2009courant}.
    
    As the main example, suppose that $\mf g$ is a quadratic Lie algebra (i.e.\ a Lie algebra with an invariant pairing) and $\mf h\subset\mf g$ is a coisotropic subalgebra (i.e.\ Lie subalgebra satisfying $\mf h^\perp\subset \mf h$). Let $H\subset G$ be a corresponding pair of Lie groups.\footnote{We will also suppose that $G$ and $H$ are connected and $H$ is a closed subgroup.} Note that $\mf g$ is in fact a Courant algebroid over a point base $M=\text{pt}$. Importantly, it can be shown \cite{li2009courant} that there exists a unique Courant algebroid pullback
     \[\begin{tikzcd}
      \mf g\times G/H\ar[r]\ar[d]&\mf g\ar[d]\\
      G/H\ar[r]&\text{pt}
    \end{tikzcd}\]
    along the trivial map $G/H\to \text{pt}$, such that the anchor map \[a\colon \mf g\times G/H\to T(G/H)\] coincides with the infinitesimal action of $G$ on the homogeneous space $G/H$. It follows easily from the transitivity of this action that the Courant algebroid $\mf g\times G/H$ is also transitive.
    
    More generally, one can start with the same data, together with a principal $G$-bundle $P\to M$ with vanishing first Pontryagin class, and then obtain particular (transitive) Courant algebroid pullbacks along the map $P/H\to P/G=M$. For more details see Example 5.10 in \cite{sv2}.

  \subsection{Poisson--Lie T-duality}  
    Suppose now we have a Courant algebroid pullback. Following \cite{sv2} we also suppose that there exists a fibrewise half-density $\tau$ on $M'$ (i.e.\ a family of half-densities defined on the fibres of the map $\pi\colon M'\to M$), which satisfies
    \begin{equation}\label{uni}
      \ms L_{\pi^*u}\tau=0,\qquad \forall u\in \Gamma(E).
    \end{equation}
    Note that this action of the Lie derivative is meaningful, since $a(\pi^*u)$ preserves the distribution $\ker\pi_*$ on $M'$. 
  
    For instance in the simple model example above there is only one fibre, namely the entire space $M'=G/H$. Hence the condition \eqref{uni} reduces to the existence of a $G$-invariant half-density on $G/H$. Such a half-density exists if and only if $\mf h$ is unimodular.
    
    Starting from a half-density $\sigma$ on $M$, we can now create a new half-density $\sigma':=\tau \pi^*\sigma$ on $M'$. The condition \eqref{uni} then ensures that Levi-Civita connections also transport nicely, namely for any $\gm$ and $\sigma$ on $E$ we have
    \begin{equation}\label{lcpull}
      LC(\gm,\sigma)\to LC(\pi^*\gm,\tau\pi^*\sigma),\qquad D\mapsto \pi^*D,
    \end{equation}
    where $\pi^*D$ is the (unique) generalised connection on $E'$ satisfying
    \[(\pi^*D)_{\pi^*u}(\pi^*v)=\pi^*(D_uv),\qquad \forall u,v\in \Gamma(E).\]
    
    Finally, we assume that $E$ satisfies the requirements of the setup of Subsection \ref{subsec:generalisation}, namely it admits a generalised metric whose $C_+$ is spin and of the required signature, and has $\on{rank}C_-\neq1$. Any pullback Courant algebroid will then automatically satisfy these conditions as well. We can then formulate the core statement of Poisson--Lie T-duality in the context of supergravity as follows:\vspace{.2cm}
    
    \noindent \emph{If the fields $(\gm, \sigma, \rho, \psi)$ satisfy the equations of motion \eqref{eqs} on $E$ then so do the fields \[(\pi^*\gm, \,\tau\pi^*\sigma,\, \tau\pi^*\rho,\, \tau\pi^*\psi)\] on $E'$. Similarly, if the former field configuration preserves some supersymmetry $\epsilon$, the latter one is supersymmetric for $\tau\pi^*\epsilon$.}\vspace{.2cm}
    
    \noindent This is an immediate consequence of \eqref{lcpull}. For instance, for any spinor half-density $\lambda$ on $M$ and a section $u\in \Gamma(E)$ we have\footnote{To see this, note that any spinor half-density can be written as $\lambda=\sigma\chi$, with $\chi$ a spinor --- we then get \[(\pi^*D)_{\pi^*u}(\tau\pi^*\lambda)=(\pi^*D)_{\pi^*u}[(\tau\pi^*\sigma)(\pi^*\chi)]=(\tau\pi^*\sigma)(\pi^*D)_{\pi^*u}(\pi^*\chi)=(\tau\pi^*\sigma)\pi^*(D_u\chi)=\tau\pi^*(D_u(\sigma\chi)).\]}
    \[(\pi^*D)_{\pi^*u}(\tau\pi^*\lambda)=\tau\pi^*(D_u\lambda).\]
    Similarly we get that the Riemann tensor is \emph{natural} w.r.t.\ Courant algebroid pullbacks, i.e.\ the Riemann tensor of the pulled-back connection is the pull-back of the original Riemann tensor. Using these facts we can then write the RHS of the equations \eqref{eqs} or of the supersymmetry variations \eqref{eq:s} for the pulled-back data as the pull-backs of the RHS of the original data; since the latter vanishes, so does the former.
    Note that (in the case of equations of motion) this result is a generalisation of the result \cite{sv2} to the full theory, including the possibility of backgrounds with nontrivial fermions. For a corresponding analysis within the superspace approach see \cite{Hassler:2023nht}, which uses the language of double field theory.

    The \emph{duality} itself then arises whenever we have two different Courant algebroid pullbacks of the form
    \begin{equation*}
      \begin{tikzcd}
      E'_1\ar[r]\ar[d]&E\ar[d] &E'_2\ar[l]\ar[d]\\
      M'_1\ar[r,"\pi_1"]&M&M'_2\ar[l,"\pi_2" above]
    \end{tikzcd}
    \end{equation*}
    Any field configuration on $E$ then gives rise to two configurations on $E_1'$ and $E_2'$, which are called \emph{Poisson--Lie T-dual}. The above analysis then implies that the configuration on $E_1'$ satisfies the equations of motion if and only if the configuration on $E_2'$ does. If $E_1'$ and $E_2'$ are both transitive, the equations of motion coincide with the usual supergravity ones.
    
    The simplest example (cf.\ \cite{sv2}) of such a duality setup arises whenever we can find two unimodular coisotropic subalgebras $\mf h_1, \mf h_2\subset \mf g$ of the same quadratic Lie algebra --- this results in dual supergravity configurations on the spacetimes $G/H_1$ and $G/H_2$. In the particular case when $\mf h_1\cap \mf h_2=0$ (which requires $\dim \mf h_1=\dim \mf h_2=\tfrac12\dim \mf g$) we have local isomorphisms $G/H_1\cong H_2$ and $G/H_2\cong H_1$, with the groups $H_1$ and $H_2$ forming a dual Poisson--Lie pair \cite{drinfeld}. This in the origin of the term Poisson--Lie T-duality.


\acknowledgments

C.S.-C.~and F.V.~are supported by an EPSRC New Investigator Award, grant number EP/X014959/1. No new data was collected or generated during the course of this research. 


\appendix
  \section{Spinors in 10 dimensions}\label{app:fierz}
    \subsection{Conventions}
      We will work in ten dimensions with metric $g$ of signature $(-,+,\dots,+)$. We set \[\epsilon_{0\dots9}=-\epsilon^{0\dots 9}=1.\] Clifford relations are
      \[\{\gamma_a,\gamma_b\}=2g_{ab}.\]
      The Majorana conjugate is defined by
      \[\bar\psi:=\psi^TC,\]
      with the charge conjugation matrix $C$ satisfying
      \[C\gamma_a C^{-1}=-\gamma_a^T,\qquad C^T=-C.\]
      In particular, if $a_1\dots a_k$ consists of $k$ terms then
      \[C\gamma_{a_1\dots a_k}C^{-1}=(-1)^{\left[\frac{k+1}2\right]} \gamma_{a_1\dots b_k}^T,\]
      implying the important flip formula
      \[\bar\psi \gamma_{a_1\dots a_k}\chi=(-1)^{\left[\frac{k+1}2\right]} \bar\chi\gamma_{a_1\dots a_k}\psi,\]
      for $\psi$, $\chi$ fermionic. We set $\gamma_*:=\gamma^0\dots\gamma^9$, so that
      \[\gamma_{a_1\dots a_k}\gamma_*=(-1)^{\left[\frac k2\right]}\tfrac1{(10-k)!}\sqrt{-g}\epsilon_{a_1\dots a_k b_1\dots b_{10-k}}\gamma^{b_1\dots b_{10-k}}.\]
      We define the positive/negative chiral Majorana spinors by $\gamma_*\psi=\psi$ and $-\psi$, respectively.
    \subsection{Gamma matrix algebra and Fierz identities}
      All the spinors appearing in this subsection will be fermionic and Majorana.
      First, denoting the chirality by $\on{ch}$, we have
      \begin{equation}\label{eq:flip}
        \tfrac1{p!}(\bar\lambda_1\gamma_{a_1\dots a_nb_1\dots b_p}\lambda_2)(\bar\lambda_3 \gamma^{b_1\dots b_p}\lambda_4)=(\on{ch}\lambda_2)(\on{ch}\lambda_4)(-1)^{1+[\frac{n}2]}\tfrac{1}{q!}(\bar\lambda_1\gamma^{c_1\dots c_q}\lambda_2)(\bar\lambda_3\gamma_{a_1\dots a_nc_1\dots c_q}\lambda_4),
      \end{equation}
      where $q:=10-(n+p)$, and so in particular
      \begin{equation}\label{eq:mirror}
        \tfrac1{(10-p)!}(\bar\lambda_1\gamma_{(10-p)}\lambda_2)(\bar\lambda_3\gamma^{(10-p)}\lambda_4)=-\tfrac1{p!}(\on{ch}\lambda_2)(\on{ch}\lambda_4)(\bar\lambda_1\gamma_{(p)}\lambda_2)(\bar\lambda_3\gamma^{(p)}\lambda_4),
      \end{equation}
      where we used the simplifying notation
      \[(\cdots) \gamma_{(p)} (\cdots) \gamma^{(p)} (\cdots) = (\cdots) \gamma_{a_1\dots a_p} (\cdots) \gamma^{a_1\dots a_p} (\cdots)\]
      
      Fierz identities follow from the basic orthonormality relation, where for any spinor matrix $M$
      \[M=\frac1{32}\sum_{p=0}^{10}\frac{(-1)^{\frac{p(p-1)}2}}{p!}\gamma_{(p)}\on{tr}(\gamma^{(p)}M).\]
      Taking any expression $\bar\lambda_1\lambda_2\bar\lambda_3\lambda_4$ we can substitute the above formula for $M=\lambda_2\bar\lambda_3$. After some flipping this results in the identity (where we stripped the first and last spinor)
      \begin{equation*}
        \lambda\bar\psi=\frac1{32}\sum_{p=0}^{10}\frac{(-1)^{p+1}}{p!}\gamma_{(p)}\psi\bar\lambda\gamma^{(p)}.
      \end{equation*}
      
      In what follows, we will use the notation where both $\psi_i$ and $\lambda_j$ are fermionic chiral Majorana spinors such that all $\psi_i$ have the same chirality, and all $\lambda_j$ have the same chirality which is opposite to the chirality of $\psi_i$.
      With this understanding the last formula and \eqref{eq:mirror} imply
      \begin{align}
        (\bar\lambda_1\psi_1)(\bar\lambda_2\psi_2)&=\tfrac1{16}(\bar\lambda_1\gamma_{(1)}\lambda_2)(\bar\psi_1\gamma^{(1)}\psi_2)+\tfrac1{96}(\bar\lambda_1\gamma_{(3)}\lambda_2)(\bar\psi_1\gamma^{(3)}\psi_2)+\tfrac1{3840}(\bar\lambda_1\gamma_{(5)}\lambda_2)(\bar\psi_1\gamma^{(5)}\psi_2).\label{eq:f2}
      \end{align}
      From this one can also derive
      \begin{align}
        (\bar\lambda_1\gamma_a\lambda_2)(\bar\lambda_3\gamma^a\lambda_4)&=\tfrac12(\bar\lambda_1\gamma_a\lambda_3)(\bar\lambda_2\gamma^a\lambda_4)+\tfrac1{24}(\bar\lambda_1\gamma_{(3)}\lambda_3)(\bar\lambda_2\gamma^{(3)}\lambda_4),\label{eq:f1}\\
        (\bar\lambda_1\gamma_a\lambda_2)(\bar\psi_1\gamma^a\psi_2)&=\tfrac58(\bar\lambda_1\psi_1)(\bar\lambda_2\psi_2)+\tfrac3{16}(\bar\lambda_1\gamma_{(2)}\psi_1)(\bar\lambda_2\gamma^{(2)}\psi_2)+\tfrac1{192}(\bar\lambda_1\gamma_{(4)}\psi_1)(\bar\lambda_2\gamma^{(4)}\psi_2),\label{eq:f3}
      \end{align}
      as well as
      \begin{equation}\label{eq:f21}
      \begin{aligned}
        (\bar\lambda\gamma^{ab}\psi_1)(\bar\psi_2\gamma_a\psi_3)&=-\tfrac7{16}(\bar\lambda\gamma^{bc}\psi_2)(\bar\psi_1\gamma_c\psi_3)-\tfrac9{16}(\bar\lambda\psi_2)(\bar\psi_1\gamma^b\psi_3)-\tfrac1{32}(\bar\lambda\gamma^{bcde}\psi_2)(\bar\psi_1\gamma_{cde}\psi_3)\\
        &\qquad-\tfrac5{32}(\bar\lambda\gamma_{cd}\psi_2)(\bar\psi_1\gamma^{bcd}\psi_3)-\tfrac1{384}(\bar\lambda\gamma_{cdef}\psi_2)(\bar\psi_1\gamma^{bcdef}\psi_3).
      \end{aligned}
      \end{equation}
      In particular \eqref{eq:f2} implies
      \begin{equation}\label{eq:fc1}
        (\bar\lambda_1\psi_1)(\bar\lambda_2\psi_2)+(\bar\lambda_1\psi_2)(\bar\lambda_2\psi_1)=\tfrac1{48}(\bar\lambda_1\gamma_{(3)}\lambda_2)(\bar\psi_1\gamma^{(3)}\psi_2).
      \end{equation}
      
    Other useful identities are \cite{Bergshoeff:1981um}
    \begin{align}
      \tfrac12(\bar\lambda\gamma^{d[ab}\lambda)\bar\lambda\gamma_d\gamma^{c]}&=(\bar\lambda\gamma^{abc}\lambda)\bar\lambda\label{eq:fcc}\\
      (\bar\lambda\gamma^{abc}\lambda)\bar\lambda\gamma_{ab}&=0\label{eq:3r2}\\
      (\bar\lambda\gamma^{abc}\lambda)\bar\lambda\gamma_{abc}&=0\label{eq:3r3}.
    \end{align}
\eqref{eq:fcc} and~\eqref{eq:3r2} can be obtained by considering the third anti-symmetric power of a chiral spinor, which is an irreducible representation of the spin group, corresponding to a two-form spinor $\psi_{[ab]}$ satisfying $\gamma^a \psi_{ab} = 0$. Setting $\chi_{abc} = (\bar\lambda\gamma_{abc}\lambda)\lambda$ we thus have that $\chi_{abc} = 3 \gamma_{[a} \psi_{bc]}$. From this follows $\psi_{ab} = \frac{1}{6} \gamma^c \chi_{abc}$ which immediately implies~\eqref{eq:3r2} and substituting this back into the previous relation gives $\chi_{abc} = \frac{1}{2} \gamma_{[a} \gamma^e \chi_{bc]e}$, which is equivalent to~\eqref{eq:fcc}. Finally, \eqref{eq:3r3} follows by multiplying \eqref{eq:3r2} by $\gamma_c$.

  \section{Elements of generalised Riemannian geometry}\label{app:gg}
    \subsection{Properties of the Riemann tensor}
      Recall that the Riemann tensor for $D\in LC(\gm,\sigma)$ was defined in \eqref{riemann}. First, we observe the following simplifying property:
      \begin{equation}\label{simplifyriemann}
        \text{if $x\in \Gamma(C_+)$ and $y\in \Gamma(C_-)$ or vice versa then } \gr(w,z,x,y)=\tfrac12 x^A y^B w^D [D_A,D_B]z_D.
      \end{equation}
      In particular
      \begin{equation}\label{zigzag}
        \text{for any $D\in LC(\gm,\sigma)$ we have }\gr_{a\beta c\delta }=0, \text{ i.e.\ }\gr(C_+,C_-,C_+,C_-)=0.
      \end{equation}
      From \eqref{simplifyriemann} is follows that for $z\in \Gamma(C_+)$ and $y\in\Gamma(C_-)$ we have
      \begin{align*}
        z^ay^{\alpha }(\gr^c{}_{ac\alpha }-\gr^{\gamma }{}_{a\gamma \alpha })&=z^ay^{\alpha }(\gr^c{}_{ac\alpha }-\gr^{\gamma }{}_{\alpha \gamma  a})=\tfrac12 y^{\alpha } [D_a,D_{\alpha }]z^a-\tfrac12z^a[D_{\alpha },D_a]y^{\alpha }\\
        &=\tfrac12[y^{\alpha }D_aD_{\alpha }z^a-y^{\alpha }D_{\alpha }D_az^a-z^aD_{\alpha }D_ay^{\alpha }+z^aD_aD_{\alpha }y^{\alpha }]\\
        &=\tfrac12[D_a(y^{\alpha }D_{\alpha }z^a)-y^{\alpha }D_{\alpha }D_az^a-D_{\alpha }(z^aD_ay^{\alpha })+z^aD_aD_{\alpha }y^{\alpha }]\\
        &=\tfrac12[\on{div} ([y,z]_+)-\ms L_y\on{div} z-\on{div}([z,y]_-)+\ms L_z\on{div} y]\\
        &=\tfrac12[\on{div} [y,z]-\ms L_y\on{div} z+\ms L_z\on{div} y]\\
        &=\tfrac14(\sigma^{-1}\ms L_{[y,z]}\sigma-\ms L_y(\sigma^{-1}\ms L_z\sigma)+\ms L_z(\sigma^{-1}\ms L_y\sigma))\\
        &=\tfrac14\sigma^{-1}(\ms L_{[y,z]}-\ms L_y\ms L_z+\ms L_z\ms L_y)\sigma=0,
      \end{align*}
      where the subscripts $\pm$ denote the projection onto $C_\pm$. In other words \cite{Siegel:1993th}
      \begin{equation}\label{contractions}
        \text{for any $D\in LC(\gm,\sigma)$ we have }\gr^c{}_{ac\alpha }=\gr^{\gamma }{}_{a\gamma \alpha }.
      \end{equation}
      
      \subsection{Lichnerowicz formula}\label{app:lich}
        Here we derive the Lichnerowicz formula for the action on spinor half-densities:
        \begin{equation}
          \di^2+D^{\alpha }D_{\alpha }=-\tfrac18\mc R.
        \end{equation}
        The proof presented here is straightforward, though rather messy --- we leave the finding of a more conceptual approach (akin to the one in ordinary geometry) open.
        
        To prove the formula, we first calculate the ingredients --- following \eqref{diracloc} we have
        \begin{align*}
          \di^2&=(\gamma^a\ms L_{e_a}-\tfrac1{12}c_{abc}\gamma^{abc})(\gamma^d\ms L_{e_d}-\tfrac1{12}c_{def}\gamma^{def})\\
          &=\ms L_{e^a}\ms L_{e_a}+\tfrac12\gamma^{ab}[\ms L_{e_a},\ms L_{e_b}]-\tfrac1{12}(a(e^a)c_{def})\gamma_a\gamma^{def}-\tfrac1{12}\{\gamma_a,\gamma^{def}\}c_{def}\ms L_{e^a}\\
          &\qquad+\tfrac1{288}c_{abc}c^{def}\{\gamma^{abc},\gamma_{def}\}\\
          &=\ms L_{e^a}\ms L_{e_a}+\tfrac12\gamma^{ab}\ms L_{[e_a,e_b]}-\tfrac1{12}(a(e_a)c_{def})\gamma^{adef}-\tfrac14(a(e^a)c_{aef})\gamma^{ef}-\tfrac12c_{aef}\gamma^{ef}\ms L_{e^a}\\
          &\qquad+\tfrac1{16}c^a{}_{bc}c_{aef}\gamma^{bcef}-\tfrac1{24}c_{abc}c^{abc}.
        \end{align*}
        When acting on half-densities we have
        \[\ms L_{[e_a,e_b]}=\ms L_{c_{abC}e^C}=c_{abC}\ms L_{e^C}+\tfrac12a(e^C)c_{abC}.\]
        Similarly, one can write the Jacobi identity as
        \[a(e_{[A})c_{BC]D}+c_{E[AB}c_{C]}{}^E{}_D-\tfrac13 a(e_D)c_{ABC}=0.\]
        This in particular implies
        \[a(e_{[A})c_{BCD]}=\tfrac34c_{[AB}{}^E c_{CD]E}.\]
        Returning back to $\di^2$ this gives
        \begin{align}\label{diracsquare}
          \di^2&=\ms L_{e^a}\ms L_{e_a}+\tfrac12c_{\alpha ef}\gamma^{ef}\ms L_{e^{\alpha }}+\tfrac14(a(e^{\alpha })c_{\alpha ef})\gamma^{ef}-\tfrac1{16}c^{\alpha }{}_{bc}c_{\alpha ef}\gamma^{bcef}-\tfrac1{24}c_{abc}c^{abc}.
        \end{align}
        Similarly, on spinor half-densities we have
        \begin{align*}
          D^{\alpha }D_{\alpha }&=D_{e^{\alpha }}D_{e_{\alpha }}-D_{D_{e^{\alpha }}e_{\alpha }}\\
          &=(\ms L_{e^{\alpha }}-\tfrac14c^{\alpha }{}_{bc}\gamma^{bc}-\tfrac12(\on{div} e^{\alpha }))(\ms L_{e_{\alpha }}-\tfrac14c_{\alpha de}\gamma^{de}-\tfrac12(\on{div} e_{\alpha }))\\
          &\qquad+(\on{div} e^{\alpha })(\ms L_{e_{\alpha }}-\tfrac14c_{\alpha de}\gamma^{de}-\tfrac12(\on{div} e_{\alpha }))\\
          &=(\ms L_{e^{\alpha }}-\tfrac14c^{\alpha }{}_{bc}\gamma^{bc}+\tfrac12(\on{div} e^{\alpha }))(\ms L_{e_{\alpha }}-\tfrac14c_{\alpha de}\gamma^{de}-\tfrac12(\on{div} e_{\alpha }))\\
          &=\ms L_{e^{\alpha }}\ms L_{e_{\alpha }}-\tfrac14(a(e^{\alpha })c_{\alpha de})\gamma^{de}-\tfrac12c_{\alpha de}\gamma^{de}\ms L_{e^{\alpha }}-\tfrac12(a(e^{\alpha })\on{div} e_{\alpha })\\
          &\qquad +\tfrac1{16}c_{\alpha bc}c^{\alpha  de}\gamma^{bc}\gamma_{de}-\tfrac14(\on{div} e^{\alpha })(\on{div} e_{\alpha })\\
          &=\ms L_{e^{\alpha }}\ms L_{e_{\alpha }}-\tfrac14(a(e^{\alpha })c_{\alpha de})\gamma^{de}-\tfrac12c_{\alpha de}\gamma^{de}\ms L_{e^{\alpha }}-\tfrac12(a(e^{\alpha })\on{div} e_{\alpha })\\
          &\qquad +\tfrac1{16}c_{\alpha bc}c^{\alpha }{}_{de}\gamma^{bcde}-\tfrac18c^{\alpha bc}c_{\alpha bc}-\tfrac14(\on{div} e^{\alpha })(\on{div} e_{\alpha }).
        \end{align*}
        Together we thus get
        \begin{align*}
          \di^2+D^{\alpha }D_{\alpha }&=\ms L_{e^A}\ms L_{e_A}-\tfrac1{24}c_{abc}c^{abc}-\tfrac12a(e^{\alpha })\on{div} e_{\alpha }-\tfrac14(\on{div} e^{\alpha })(\on{div} e_{\alpha })-\tfrac18c^{\alpha bc}c_{\alpha bc}.
        \end{align*}
        Using
        \begin{align*}
          \ms L_{e^a}\ms L_{e_a}\sigma&=\tfrac12\ms L_{e^a}[\sigma(\sigma^{-2}\ms L_{e_a}\sigma^2)]=\tfrac12(\ms L_{e^a}\sigma)\on{div} e_a+\tfrac12\sigma a(e^a)\on{div} e_a\\
          &=\sigma[\tfrac14(\on{div} e^a)(\on{div} e_a)+\tfrac12a(e^a)\on{div} e_a]
        \end{align*}
        and the analogous formula for $C_-$, we finally obtain
        \[\di^2+D^{\alpha }D_{\alpha }=\tfrac14(\on{div} e^a)(\on{div} e_a)+\tfrac12a(e^a)\on{div} e_a-\tfrac1{24}c_{abc}c^{abc}-\tfrac18c^{\alpha bc}c_{\alpha bc}\stackrel{\eqref{scalarexplicit}}=-\tfrac18\gr.\]
        
      \subsection{The other formula}\label{app:lichric}
        It is much simpler to prove that on spinor half-densities one has
        \begin{equation}
          [\di, D_{\alpha }]=\tfrac14\gr_{a\alpha }\gamma^a.
        \end{equation}
      Namely, rewriting \eqref{simplifyriemann} as
      $[D_a,D_{\alpha }]^A{}_B=2\gr_{a\alpha }{}^A{}_B$,
      it follows that
      \[[\gamma^aD_a, D_{\alpha }]\epsilon=\tfrac12\gr_{a\alpha cd}\gamma^a\gamma^{cd}\epsilon\stackrel{\eqref{riemsym}}=\gr^c{}_{\alpha cd}\gamma^{d}\epsilon\stackrel{\eqref{contractions}}=\tfrac14\gr_{a\alpha }\gamma^a\epsilon.\]
        
      \subsection{Generating Dirac operator}\label{app:diff}
        Note that Sections \ref{subsec:con} and \ref{subsec:curv} apply also to the more general context where $E$ is any Courant algebroid, $\gm$ is any endomorphism with $\gm^T=\gm$ and $\gm^2=\on{id}$, and $\sigma$ an everywhere non-vanishing half-density, provided
        \[\on{rank}C_+\neq 1\qquad \text{and}\qquad\on{rank}C_-\neq1,\] since otherwise the space $LC(\gm,\sigma)$ may be empty (cf.\ \cite{Streets:2024rfo}).
        
        A particularly important special case is $\gm=\on{id}$ in which we have $C_+=E$ and $C_-=0$, and the Dirac operator becomes the \emph{generating Dirac operator} $\di_{\text{gen}}$ of Alekseev--Xu \cite{AXu,let}. In this case the Lichnerowicz formula \eqref{lich} gives
        \[\di^2_{\text{gen}}=-\tfrac18 \gr=-\tfrac14\gr^{AB}{}_{AB}\in C^\infty(M)\qquad \text{for any }D\in LC(\on{id},\sigma).\]
        Using the fact that for any $\gm$ we have $LC(\gm,\sigma)\subset LC(\on{id},\sigma)$, we get
        \begin{equation}\label{contractions2}
          \gr^{ab}{}_{ab}+\gr^{\alpha \beta }{}_{\alpha \beta }\stackrel{\eqref{zigzag}}=\gr^{AB}{}_{AB}=-4\di^2_{\text{gen}}\qquad \text{for any }D\in LC(\gm,\sigma).
        \end{equation}
        Since $\di_{\text{gen}}$ is independent of both $\sigma$ (cf.\ \eqref{diracloc}) and $\gm$, so is the sum $\gr^{ab}{}_{ab}+\gr^{\alpha \beta }{}_{\alpha \beta }$.
        
        For completeness we note that in the case \eqref{bkt} the formula \eqref{diracsquare} implies
        \[\di^2_{\text{gen}}=-\tfrac1{24}f_{ijk}f^{ijk}=\text{const},\]
        where $f_{ijk}$ are the structure coefficients of $\mf g$.
        
      \subsection{Variations of the kinetic operators}\label{subsec:vars}
        The variation of $\Gamma_{ABC}$ and of the curvature tensors under the change of $\gm$ and $\sigma$ was calculated in \cite{Streets:2024rfo}. Here we will only need\footnote{Note that $(\delta \Gamma)_{ABC}:=\Gamma_{ABC}^{\text{new}}-\Gamma_{ABC}^{\text{old}}$, with both expressions evaluated in the original frame.}
        \[(\delta\Gamma)_{[abc]}=0,\quad (\delta\Gamma)_{a\alpha \gamma }=D_{[\alpha }\delta\gm_{\gamma ]a},\quad (\delta\Gamma)_{\alpha bc}=-D_{[b}\delta\gm_{c]\alpha },\quad (\delta\Gamma)^{\gamma }{}_{\gamma \alpha }=-\tfrac12D_a\delta\gm_{\alpha }{}^a+2D_{\alpha }\tfrac{\delta\sigma}\sigma\]
        as well as
        \begin{equation}\label{varr}
          \int_M(\delta\!\gr)\sigma^2=\int_M\gr_{a\alpha }(\delta\gm)^{a\alpha }\sigma^2.
        \end{equation}
      Note that both $\delta\gm_{ab}$ and $\delta\gm_{\alpha\beta}$ always vanish as a consequence of \[0=\delta(\on{id})=\delta\gm^2=(\delta\gm) \gm+\gm(\delta\gm).\]

        We now wish to show the following variations:\footnote{Note that it is not completely obvious (but it is still true) that the RHS of these expressions are independent of the choice of the representative $D$ and that the first two are also independent of $\sigma$.}
        \begin{align*}
          (\delta\di)\rho&=\tfrac12\delta \gm^{\alpha }{}_a\gamma^aD_{\alpha }\rho+\tfrac14 (D_{\alpha } \delta\gm^{\alpha }{}_a)\gamma^a\rho,\\
          (\delta\di)\psi^{\alpha }&=\tfrac12\delta \gm^{\gamma }{}_a\gamma^aD_{\gamma }\psi^{\alpha }+\tfrac14 (D_{\gamma } \delta\gm^{\gamma }{}_a)\gamma^a\psi^{\alpha }+(D^{[\alpha }\delta\gm^{\gamma ]}{}_a)\gamma^a\psi_{\gamma },\\
          (\delta D_{\alpha })\rho&=-\tfrac12\delta\gm_{\alpha }{}^aD_a\rho-\tfrac14(D_b\delta\gm_{\alpha c})\gamma^{bc}\rho-(D_{\alpha }\tfrac{\delta\sigma}\sigma)\rho,
        \end{align*}
        which then yield
        \begin{align*}
          \int_M\bar\rho(\delta\di)\rho&=\int_M\tfrac12\delta\gm^{\alpha }{}_a(\bar\rho\gamma^aD_{\alpha }\rho),\\
          \int_M\bar\psi_{\alpha }(\delta\di)\psi^{\alpha }&=\int_M\delta\gm_{\gamma a}(\tfrac12\bar\psi_{\alpha }\gamma^aD^{\gamma }\psi^{\alpha }+\bar\psi^{\gamma }\gamma^aD^{\alpha }\psi_{\alpha }-\bar\psi_{\alpha }\gamma^aD^{\alpha }\psi^{\gamma }),\\
          \int_M\bar\psi^{\alpha }(\delta D_{\alpha })\rho&=\int_M\delta \gm_{\alpha a}(-\tfrac12\bar\psi^{\alpha }D^a\rho+\tfrac14\bar\rho\gamma^{ab}D_b\psi^{\alpha }-\tfrac14\bar\psi^{\alpha }\gamma^{ab}D_b\rho)+\tfrac{\delta\sigma}\sigma(\bar\psi^{\alpha }D_{\alpha }\rho+\bar\rho D_{\alpha }\psi^{\alpha }).
        \end{align*}

        To prove the variation formulas we note that we can take \[\delta e_a=\tfrac12 \delta\gm^{\alpha }{}_ae_{\alpha },\qquad \delta e_{\alpha }=-\tfrac12\delta\gm^a{}_{\alpha }e_a,\]
        and that the connection coefficients transform under the change of basis $\delta e_A=M^B{}_A e_B$ as
        \[\delta(\Gamma_{ABC})=M^D{}_A \Gamma_{D BC}+M^D{}_B \Gamma_{AD C}+M^D{}_C \Gamma_{ABD}+a(e_A)M_{BC}.\]
        We then directly calculate
        \begin{align*}
          (\delta\di)\rho&=\delta(\gamma^a\ms L_{e_a}+\tfrac14\Gamma_{abc}\gamma^{abc})\rho\\
          &=\gamma^a\ms L_{\frac12\delta\gm^{\alpha }{}_ae_{\alpha }}\rho+\tfrac14((\delta\Gamma)_{abc}+\tfrac12\delta\gm^{\alpha }{}_a\Gamma_{\alpha bc}+\tfrac12\delta\gm^{\beta }{}_b\Gamma_{a\beta c}+\tfrac12\delta\gm^{\gamma }{}_c\Gamma_{ab\gamma })\gamma^{abc}\rho\\
          &=\gamma^a\ms L_{\frac12\delta\gm^{\alpha }{}_ae_{\alpha }}\rho+\tfrac18\delta\gm^{\alpha }{}_a\Gamma_{\alpha bc}\gamma^{abc}\rho\\
          &=\tfrac12\gamma^a\delta\gm^{\alpha }{}_a\ms L_{e_{\alpha }}\rho+\tfrac14\gamma^a(a(e_{\alpha })\delta\gm^{\alpha }{}_a)\rho+(\tfrac18\delta\gm^{\alpha }{}_a\Gamma_{\alpha bc}\gamma^a\gamma^{bc}\rho-\tfrac14\delta\gm^{\alpha }{}_a\Gamma_{\alpha }{}^{a}{}_c\gamma^c\rho)\\
          &\qquad+(\tfrac14\gamma^a\delta\gm^{\alpha }{}_a\Gamma^{\gamma }{}_{\gamma \alpha }\rho-\tfrac14\gamma^a\delta\gm^{\alpha }{}_a\Gamma^{\gamma }{}_{\gamma \alpha }\rho)\\
          &=\tfrac12\gamma^a\delta\gm^{\alpha }{}_aD_{\alpha }\rho+\tfrac14\gamma^a(D_{\alpha }\delta\gm^{\alpha }{}_a)\rho,\\
          (\delta\di)\psi^{\alpha }&=\gamma^a\ms L_{\delta e_a}\psi^{\alpha }+\tfrac14\delta(\Gamma_{abc})\gamma^{abc}\psi^{\alpha }+\delta(\Gamma_{a}{}^{\alpha }{}_{\gamma })\gamma^a\psi^{\gamma }\\
          &=\gamma^a\ms L_{\frac12\delta\gm^{\gamma }{}_ae_{\gamma }}\psi^{\alpha }+\tfrac18\delta\gm^{\gamma }{}_a\Gamma_{\gamma bc}\gamma^{abc}\psi^{\alpha }+(D^{[\alpha }\delta\gm^{\gamma ]}{}_a+\tfrac12\delta\gm^{\beta }{}_a\Gamma_{\beta }{}^{\alpha }{}_{\gamma })\gamma^a\psi_{\gamma }\\
          &=\tfrac12\gamma^a\delta\gm^{\gamma }{}_a\ms L_{e_{\gamma }}\psi^{\alpha }+\tfrac14\gamma^a(a(e_{\gamma })\delta\gm^{\gamma }{}_a)\psi^{\alpha }+(\tfrac18\delta\gm^{\gamma }{}_a\Gamma_{\gamma bc}\gamma^a\gamma^{bc}\psi^{\alpha }-\tfrac14\delta\gm^{\gamma }{}_a\Gamma_{\gamma }{}^{a}{}_c\gamma^c\psi^{\alpha })\\
          &\qquad+D^{[\alpha }\delta\gm^{\gamma ]}{}_a\gamma^a\psi_{\gamma }+\tfrac12\delta\gm^{\beta }{}_a\Gamma_{\beta }{}^{\alpha }{}_{\gamma }\gamma^a\psi_{\gamma }+(\tfrac14\gamma^a\delta\gm^{\beta }{}_a\Gamma^{\gamma }{}_{\gamma \beta }\psi^{\alpha }-\tfrac14\gamma^a\delta\gm^{\beta }{}_a\Gamma^{\gamma }{}_{\gamma \beta }\psi^{\alpha })\\
          &=\tfrac12\gamma^a\delta\gm^{\gamma }{}_aD_{\gamma }\psi^{\alpha }+\tfrac14\gamma^a(D_{\gamma }\delta\gm^{\gamma }{}_a)\psi^{\alpha }+D^{[\alpha }\delta\gm^{\gamma ]}{}_a\gamma^a\psi_{\gamma },\\
          (\delta D_{\alpha })\rho&=\ms L_{\delta e_{\alpha }}\rho+\tfrac14\delta(\Gamma_{\alpha bc})\gamma^{bc}\rho-\tfrac12\delta(\Gamma^{\gamma }{}_{\gamma \alpha })\rho\\
          &=-\tfrac12\delta\gm^a{}_{\alpha }\ms L_{e_a}\rho-\tfrac14(a(e_a)\delta\gm^a{}_{\alpha })\rho-\tfrac14(D_{b}\delta\gm_{c\alpha })\gamma^{bc}\rho-\tfrac18\delta\gm^a{}_{\alpha }\Gamma_{abc}\gamma^{bc}\rho\\
          &\qquad+\tfrac14(D_a\delta\gm_{\alpha }{}^a)\rho-(D_{\alpha }\tfrac{\delta\sigma}\sigma)\rho+\tfrac14\delta\gm^{\gamma }{}_a\Gamma^a{}_{\gamma \alpha }\rho\\
          &=-\tfrac12\delta\gm^a{}_{\alpha }\ms L_{e_a}\rho-\tfrac14(a(e_a)\delta\gm^a{}_{\alpha })\rho-\tfrac14(D_{b}\delta\gm_{c\alpha })\gamma^{bc}\rho-\tfrac18\delta\gm^a{}_{\alpha }\Gamma_{abc}\gamma^{bc}\rho\\
          &\qquad+\tfrac14\delta\gm_{\alpha }{}^c\Gamma_a{}^a{}_{c}\rho-(D_{\alpha }\tfrac{\delta\sigma}\sigma)\rho\\
          &=-\tfrac12\delta\gm^a{}_{\alpha }D_{a}\rho-\tfrac14(D_{b}\delta\gm_{c\alpha })\gamma^{bc}\rho-(D_{\alpha }\tfrac{\delta\sigma}\sigma)\rho.
        \end{align*}

  \section{Unpacking the generalised geometry} \label{app:unpack}
    \subsection{Calculating the brackets}
      We consider the local model
      \[E=TM\oplus T^*M\oplus (\mf g\times M),\]
      with the structure
      \begin{equation*}
        \begin{gathered}
          a(x+\alpha+s):=x,\qquad \langle x+\alpha+s,y+\beta+t\rangle:=\alpha(y)+\beta(x)+\tr st\\
          [x+\alpha+s,y+\beta+t]:=L_xy+(L_x\beta-i_yd\alpha+\tr t\,ds)+(L_xt-L_ys+[s,t]_\mf g)
        \end{gathered}
      \end{equation*}
      and a generalised metric given by
      \begin{align*}
        E&=C_+\oplus C_-=C_+\oplus (C_-'\oplus C_-'')\\
        C_+&=\{x+(i_xg+i_xB-\tfrac12\tr A \,i_xA)+i_xA\mid x\in TM\}\\
        C'_-&=\{x+(-i_xg+i_xB-\tfrac12\tr A \,i_xA)+i_xA\mid x\in TM\}\\
        C''_-&=\{0-\tr tA+t\mid t\in \mf g\times M\}.
      \end{align*}
      As the first step, we will identify $E$ with $TM\oplus TM\oplus (\mf g\times M)$ via the bundle isomorphisms
      \[j_+:=(a|_{C_+})^{-1}\colon TM\to C_+,\qquad j_-:=(a|_{C_-'})^{-1}\colon TM\to C_-',\]
      \[j_\pm x=x+(\pm i_xg+i_xB-\tfrac12\tr A \,i_xA)+i_xA.\]
      and
      \[j_\mf g\colon M\times \mf g\to C_+'',\qquad j_\mf gt=-\tr t A+t.\]
      A straightforward calculation then gives
      \begin{align*}
        [j_\pm x,j_\pm y]&=j_\pm[x,y]\pm 2g(\nabla x,y)+i_yi_xdB+\tfrac12i_yi_x\tr (A\wedge dA)-\tr (A\,i_yi_xdA)+i_yi_xF\\
        [j_\pm x,j_\mf gt]&=j_\mf g(i_x\nabla_{\!A}t)-\tr [t(i_XF)],\\
        [j_\mf gs,j_\mf gt]&=j_\mf g([s,t]_\mf g)+\tr (t \nabla_{\!A}s),
      \end{align*}
      and the subsequent
      \begin{align*}
        \la [j_\pm x,j_\pm y],j_\pm z\ra&=\pm2g([x,y],z)\pm2g(\nabla_zx,y)+i_zi_yi_xH,\\
        \la [j_\pm x,j_\pm y],j_\mp z\ra&=\pm2g(\nabla_zx,y)+i_zi_yi_xH,
      \end{align*}
      where $\nabla$ is the (ordinary) Levi-Civita connection, $\nabla_{\!A}t:=dt+[A,t]_\mf g$, and
      \begin{equation}\label{conv}
        F:=dA+\tfrac12[A,A]_\mf g,\qquad H:=dB+\tfrac12\on{cs}(A),\qquad \on{cs}(A):=\tr(A\wedge dA)+\tfrac13\tr(A\wedge [A,A]_\mf g).
      \end{equation}
      
    \subsection{Structure coefficients via a normal frame}\label{subsec:strcoeff}
      We will now use the standard argument using a normal frame (cf.\ \cite{sv2}). Let us pick --- around any point $p\in M$ --- a local frame $E_a$ of $TM$ satisfying $g(E_a,E_b)=g_{ab}=\text{const}$ (in the entire neighbourhood) and $\bar\Gamma_{abc}=0$ at $p$, where $\bar\Gamma$ are the usual connection coefficients (in particular we have $[E_a,E_b]=0$ at $p$). We also choose a basis $E_i$ of the Lie algebra $\mf g$. We then define the following frame of $C_+$:
      \[e_a:=\tfrac1{\sqrt2}(j_+E_a),\]
      which gives $\la e_a,e_b\ra=g_{ab}$, and so in particular also $e^a=g^{ab}e_b$.
      We also define the frames
      \[e_{\dot a}:=\tfrac1{\sqrt2}(j_-E_a),\qquad e_i:=j_\mf gE_i\]
      of $C_-'$ and $C_-''$, respectively. (In particular we now have a further splitting of the frame $e_\alpha$ into $e_{\dot a}$ and $e_i$.)
      
      Next, using the fact that the images of $j_+$, $j_-$, and $j_\mf g$ are mutually orthogonal, we calculate the needed ingredients at the point $p$:
      \[c_{abc}=\la [e_a,e_b],e_c\ra=\tfrac1{2\sqrt2}\la [j_+E_a,j_+E_b],j_+E_c\ra=\tfrac1{2\sqrt2}H_{abc},\]
      and similarly
      \[c_{ab\dot c}=\tfrac1{2\sqrt2}H_{abc},\quad c_{abi}=\tfrac12(F_{ab})_i,\quad c_{a\dot b\dot c}=\tfrac1{2\sqrt2}H_{abc},\quad c_{a\dot b i}=\tfrac12(F_{ab})_i,\quad c_{aij}=\tfrac1{\sqrt2}(A_a)^kf_{kij},\]
      where $f_{ijk}$ are the structure coefficients of $\mf g$.
      Recalling the relation \eqref{eq:dil}, locally (i.e.\ not just at $p$) we have
      \[\on{div} e_a=\on{div} e_{\dot a}=\tfrac1{\sqrt2}(-2{E_a}\varphi+\bar\Gamma^c{}_{ca})\]
      and so at $p$ we get $\on{div} e_a=\on{div} e_{\dot a}=-\sqrt2{E_a}\varphi$.
        
    \subsection{Scalar curvature}
      At $p$ we then have
      \begin{align*}
        \gr^{ab}{}_{ab}&=-(\on{div} e^a)(\on{div} e_a)-2a(e^a)\on{div} e_a-\tfrac13c_{abc}c^{abc}+\tfrac12c_{abC}c^{abC}\\
        &=-2(E^a\varphi)(E_a\varphi)+2E^aE_a\varphi-\bar\Gamma^c{}_{ca,}{}^a-\tfrac13\la [e_a,e_b],e_c\ra\la [e^a,e^b],e^c\ra+\tfrac12\la [e_a,e_b],[e^a,e^b]\ra\\
        &=-2(E^a\varphi)(E_a\varphi)+2E^aE_a\varphi-\bar\Gamma^c{}_{ca,}{}^a-\tfrac1{24}H_{abc}H^{abc}+\tfrac18\tr F_{ab}F^{ab},
      \end{align*}
      and so
      \[\gr =R+4\Delta \varphi-4g(\nabla \varphi,\nabla\varphi)-\tfrac1{12}H_{abc}H^{abc}+\tfrac14\tr F_{ab}F^{ab},\]
      where $R$ is the usual scalar curvature for $g$.
      
  \subsection{Fermionic kinetic terms}
  Recall that the fields are decomposed as
  \[\rho=\sqrt[4]2\sigma\uprho,\qquad \psi^{\dot a}=\sqrt[4]2\sigma\uppsi^a,\qquad \psi^i=\tfrac1{\sqrt[4]2}\sigma\upchi^i.\]
  We first calculate (noting that $\psi_{\dot a}=-\sqrt[4]2\sigma\uppsi_a$ due to the minus sign in \eqref{notisometry}) that at the point $p$
  \begin{align*}
    \di\rho&\stackrel{\eqref{diracloc}}=\sqrt[4]2[\gamma^a\ms L_{e_a}(\sigma\uprho)-\tfrac1{12}\sigma c_{abc}\gamma^{abc}\uprho]=\sqrt[4]2[\tfrac12(\on{div}e_a)\sigma\gamma^a\uprho+\sigma\gamma^aa(e_a)\uprho-\tfrac1{12}\sigma c_{abc}\gamma^{abc}\uprho]\\
    &=\tfrac1{\sqrt[4]2}\sigma[-(E_a\varphi)\gamma^a\uprho+\gamma^aE_a\uprho-\tfrac1{24}H_{abc}\gamma^{abc}\uprho]=\tfrac1{\sqrt[4]2}\sigma(-(\nabla_a\varphi)\gamma^a\uprho+\slashed\nabla\uprho-\tfrac14\slashed H\uprho)\\
    \di\psi_{\dot a}&\stackrel{\eqref{diracloc2}}=-\tfrac1{\sqrt[4]2}\sigma(-(\nabla_c\varphi)\gamma^c\uppsi_a+\slashed\nabla\uppsi_a-\tfrac14\slashed H\uppsi_a+\sqrt2c_{b\dot a\dot c}\gamma^b\uppsi^c+c_{b\dot ai}\gamma^b\upchi^i)\\
    &=\tfrac1{\sqrt[4]2}\sigma[(\nabla_c\varphi)\gamma^c\uppsi_a-\slashed\nabla\uppsi_a+\tfrac14\slashed H\uppsi_a+\tfrac12 H_{abc}\gamma^b\uppsi^c+\tfrac12\tr F_{ab}\gamma^b\upchi]\\
    \di\psi_i&\stackrel{\eqref{diracloc2}}=\tfrac{\sqrt[4]2}{2}\sigma[-(\nabla_a\varphi)\gamma^a\upchi_i+\slashed\nabla\upchi_i-\tfrac14\slashed H\upchi_i-2c_{bi\dot c}\gamma^b\uppsi^c-\sqrt2c_{bij}\gamma^b\upchi^j]\\
    &=\tfrac{\sqrt[4]2}2\sigma[-(\nabla_a\varphi)\gamma^a\upchi_i+\slashed\nabla_{\hspace{-1mm}A}\upchi_i-\tfrac14\slashed H\upchi_i-(F_{ab})_i \gamma^b\uppsi^a]\\
    D_{\dot a}\rho&\stackrel{\eqref{derivloc}}=\sqrt[4]2[\ms L_{e_{\dot a}}(\sigma\uprho)-\tfrac14c_{\dot abc}\gamma^{bc}(\sigma\uprho)-\tfrac12(\on{div} e_{\dot a})(\sigma\uprho)]\\
    &=\tfrac1{\sqrt[4]2}\sigma[-(E_a\varphi)\uprho+E_a\uprho-\tfrac18H_{abc}\gamma^{bc}\uprho+(E_a\varphi)\uprho]=\tfrac1{\sqrt[4]2}\sigma[\nabla_a\uprho-\tfrac18H_{abc}\gamma^{bc}\uprho]\\
    D_i\rho&\stackrel{\eqref{derivloc}}=-\tfrac{\sqrt[4]2}{4}\sigma c_{ibc}\gamma^{bc}\uprho=-\tfrac{\sqrt[4]2}8\sigma(F_{bc})_i\gamma^{bc}\uprho=-\tfrac{\sqrt[4]2}4\sigma\slashed F_{\!i}\uprho
  \end{align*}
  where
  \begin{equation}\label{conva}
    \slashed\nabla_{\hspace{-1mm}A}\upchi:=\slashed\nabla \upchi+\gamma^a[A_a,\upchi]_\mf g.
  \end{equation}
  For the kinetic terms we then have
  \begin{align*}
    \bar\rho \di\rho&=\sigma^2(\uprho\slashed\nabla\uprho-\tfrac14\bar\uprho\slashed H\uprho)\\
    \bar\psi^{\alpha }\di\psi_{\alpha }&=\sigma^2[-\bar\uppsi^a\slashed\nabla\uppsi_a+\tfrac14\bar\uppsi^a\slashed H\uppsi_a+\tfrac12\tr(\bar\upchi\slashed\nabla_{\hspace{-1mm}A}\upchi-\tfrac14\bar\upchi\slashed H\upchi)+\tfrac12 H_{abc}\bar\uppsi^a \gamma^b\uppsi^c+\tr F_{ab}\bar\uppsi^a \gamma^b\upchi]\\
    \bar\psi^{\alpha }D_{\alpha }\rho&=\sigma^2(\bar\uppsi^a\nabla_a\uprho-\tfrac18\bar\uppsi^aH_{abc}\gamma^{bc}\uprho-\tfrac14\tr\bar\upchi\slashed F\uprho).
  \end{align*}
  Together this gives $S=\int_M\sqrt{|g|}e^{-2\varphi}L$ with
  \begin{align*}
    L&=R+4|\nabla \varphi|^2-\tfrac1{12}H_{abc}H^{abc}+\tfrac14\tr F_{ab}F^{ab}-\bar\uppsi^a\slashed\nabla\uppsi_a+\uprho\slashed\nabla\uprho+\tfrac12\tr\bar\upchi\slashed\nabla_{\hspace{-1mm}A}\upchi-2\bar\uppsi^a\nabla_a\uprho\\
    &\qquad +\tfrac14\bar\uppsi^a\slashed H\uppsi_a-\tfrac14\bar\uprho\slashed H\uprho-\tfrac18\tr\bar\upchi\slashed H\upchi+\tfrac12 H_{abc}\bar\uppsi^a \gamma^b\uppsi^c+\tfrac14\bar\uppsi^aH_{abc}\gamma^{bc}\uprho\\
    &\qquad +\tfrac12\tr\bar\upchi\slashed F\uprho+\tr F_{ab}\bar\uppsi^a \gamma^b\upchi+\tfrac1{384}(\bar\uppsi_{a}\gamma_{bcd}\uppsi^{a})(\bar\uprho\gamma^{bcd}\uprho)-\tfrac1{768}(\bar\uprho\gamma^{bcd}\uprho)\tr(\bar\upchi\gamma_{bcd}\upchi)\\
    &\qquad -\tfrac1{192}(\bar\uppsi_{a}\gamma_{cde}\uppsi^{a})(\bar\uppsi_{b}\gamma^{cde}\uppsi^{b})+\tfrac1{192}(\bar\uppsi_{a}\gamma_{cde}\uppsi^{a})\tr(\bar\upchi\gamma^{cde}\upchi)-\tfrac1{768}\tr(\bar\upchi\gamma_{abc}\upchi)\tr(\bar\upchi\gamma^{abc}\upchi),
  \end{align*}
  where we used $\int_M\sqrt{|g|}e^{-2\varphi}\Delta\varphi=2\int_M\sqrt{|g|}e^{-2\varphi}|\nabla \varphi|^2$. Switching to the more standard $\mu,\nu,\dots$ spacetime indices we then obtain \eqref{eq:origa}.

    \subsection{Variation of the generalised metric}
      An infinitesimal variation $\delta\gm$ of a generalised metric $\gm$ can be equivalently described via a map $\tau\colon C_+\to C_-$ (the graph of $\epsilon \tau$, with $\epsilon$ a small parameter, corresponds the the new deformed generalised metric). 
      To express the latter in terms of $\delta g$, $\delta B$, and $\delta A$ via the correspondence \eqref{cplusparam} we start with $j_+x\in C_+$ and then identify $\tau (j_+x)$ as the unique element in $C_-$ for which 
      \[j_+x+\epsilon\tau (j_+x)\in\on{Im}j_+^{\text{new}},\]
      up to order $\epsilon$, where 
      \[j_+^{\text{new}}y=y+[i_y(g+\epsilon\delta g)+i_y(B+\epsilon\delta B)-\tfrac12\tr (A+\epsilon\delta A)i_y(A+\epsilon\delta A)]+i_y(A+\epsilon\delta A).\]
    A quick calculation then reveals
    \[\tau(j_+x)=j_-(-\tfrac12g^{-1} i_x(\delta g+\delta B+\tfrac12\tr \delta A\wedge A))+j_\mf g(i_x\delta A).\]
    
    Using the frame from Subsection \ref{subsec:strcoeff} and the relation
      \[\delta\gm_{a\alpha}=\la(\delta\gm)e_a,e_\alpha\ra=2\la \tau e_a,e_\alpha\ra\]
    we obtain
    \[\delta \gm_{a\dot c}=(\delta g+\delta B+\tfrac12\tr \delta A\wedge A)_{ac},\qquad \delta \gm_{ai}=\sqrt2(\delta A)_{ai}.\]
    The supersymmetry variations \eqref{eq:origs} then follow directly (using $\epsilon=-\tfrac2{\sqrt[4]2}\sigma\upepsilon$).


\bibliographystyle{JHEP}
\bibliography{citations}


\end{document}